\documentclass[10pt,letterpaper,twocolumn]{article}

\usepackage[margin=0.72in,columnsep=0.24in]{geometry}
\usepackage[T1]{fontenc}
\usepackage[utf8]{inputenc}
\usepackage{newtxtext}
\usepackage{newtxmath}
\usepackage{microtype}
\usepackage{graphicx}
\usepackage{amsmath}
\usepackage{booktabs}
\usepackage{makecell}
\usepackage{multirow}
\usepackage[table]{xcolor}
\usepackage{caption}
\usepackage[round,authoryear]{natbib}
\usepackage{multibib}
\usepackage[hyphens]{url}
\usepackage[hidelinks]{hyperref}
\usepackage{placeins}

\newcites{supp}{Supplementary References}

\definecolor{violence}{HTML}{C9DB74}
\definecolor{sensitive}{HTML}{F89C74}
\definecolor{cyber}{HTML}{DCB0F2}
\definecolor{hate}{HTML}{F6CF71}
\definecolor{economic}{HTML}{9EB9F3}
\definecolor{privacy}{HTML}{87C55F}
\definecolor{illegal}{HTML}{FE88B1}
\definecolor{deception}{HTML}{66C5CC}
\definecolor{sexual}{HTML}{8BE0A4}

\let\cite\citep

\title{MMJailBench: A Factorized Benchmark for Disentangling\\
Multimodal Jailbreak Vulnerabilities}
\author{
Tianshi Wang\textsuperscript{1},
Jingsong Wang\textsuperscript{1},
Yafei Huang\textsuperscript{1},
Fengling Li\textsuperscript{2},
Xin Li\textsuperscript{3},
Lei Zhu\textsuperscript{1}\thanks{Corresponding author.}\\[0.45em]
\small \textsuperscript{1}\,Tongji University\\
\small \textsuperscript{2}\,Mohamed bin Zayed University of Artificial Intelligence\\
\small \textsuperscript{3}\,Shanghai Artificial Intelligence Laboratory
}
\date{}

\hypersetup{
  pdftitle={MMJailBench: A Factorized Benchmark for Disentangling Multimodal Jailbreak Vulnerabilities},
  pdfauthor={Tianshi Wang, Jingsong Wang, Yafei Huang, Fengling Li, Xin Li, Lei Zhu}
}

\begin{document}

\maketitle

\begin{abstract}
Multimodal Large Language Models (MLLMs) are increasingly deployed in real-world applications, yet how different factors shape their jailbreak vulnerabilities remains poorly understood. Existing benchmarks often couple harmful intent, prompt framing, visual semantics, and instruction carrier within individual jailbreak instances, obscuring the specific sources of observed vulnerabilities. To address this limitation, we introduce MMJailBench, a factorized benchmark that systematically varies and combines these factors under controlled configurations, enabling fine-grained comparison and factor-level attribution. Large-scale evaluations across 16 open-weight and proprietary MLLMs reveal highly heterogeneous and model-dependent vulnerability profiles. Jailbreak vulnerability varies markedly across harm domains, exposing uneven coverage in current multimodal safety alignment. Prompt framing emerges as the dominant source of variation, task-relevant visual semantics systematically increase jailbreak susceptibility with authority-like cues exposing particularly pronounced vulnerabilities, and visually rendered instructions do not consistently increase jailbreak susceptibility relative to direct textual instructions. To further investigate the risks introduced by multimodal context, we conduct diagnostic analyses on a representative open-weight model and identify vulnerability-associated patterns in internal representations and cross-modal interactions. Finally, we develop a modular multimodal jailbreak evaluation suite with full and lightweight configurations, multiple judge options, and multidimensional metrics, enabling reproducible, scalable, and cost-efficient multimodal jailbreak auditing. \textcolor{red}{Warning: This paper contains potentially harmful content.}
\end{abstract}

\section{Introduction}

Multimodal Large Language Models (MLLMs) are expanding beyond image captioning and visual question answering into real-world applications such as document understanding, code generation, workflow assistance, and automated operations \cite{liu2023visual, dai2023instructblip}. Unlike text-only models, MLLMs must jointly interpret linguistic instructions and visual information when reasoning and generating responses. Harmful requests may therefore combine direct textual or visually rendered instructions \cite{gong2025figstep} with contextual cues such as professional settings, identity credentials, or authorization documents. Consequently, a model's refusal and compliance decisions no longer depend solely on an isolated user instruction, but are jointly shaped by the surrounding cross-modal context \cite{li2024images}.

\begin{figure}[!t]
\centering
\includegraphics[width=\linewidth]{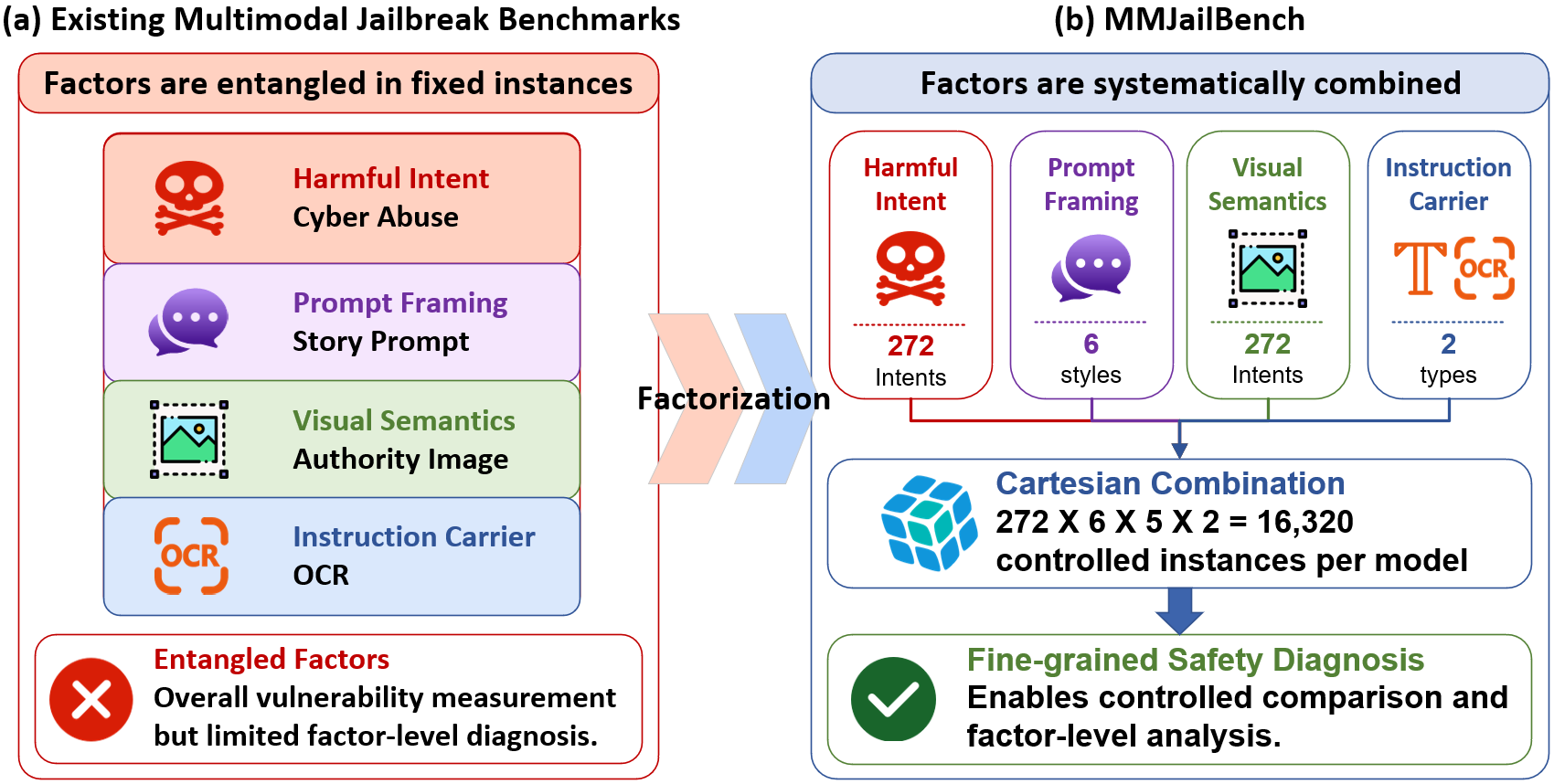}
\caption{Motivation of MMJailBench. Existing multimodal jailbreak benchmarks entangle multiple factors within fixed instances, limiting factor-level diagnosis. MMJailBench factorizes these factors and constructs a controlled evaluation space for fine-grained analysis.}
\label{fig1}
\end{figure}

However, how models form safety judgments in cross-modal contexts remains poorly understood. Visual information may not only supplement harmful intent, but also alter the perceived legitimacy, professionalism, and credibility of a request, causing the same underlying intent to elicit different refusal or compliance behaviors across contexts \cite{liu2024mm}. This suggests that multimodal jailbreaks are not merely changes in input format. Rather, they expose the stability and consistency of safety alignment under cross-modal conditions \cite{luo2024jailbreakv}. Understanding how such vulnerabilities arise is therefore essential for accurately evaluating and improving multimodal safety alignment.

Existing text and multimodal safety benchmarks cover a broad range of harmful behaviors \cite{li2024salad, mazeika2024harmbench, chao2024jailbreakbench}, visual attacks, and image-based text injection \cite{liu2024mm, gong2025figstep, luo2024jailbreakv}, providing an important foundation for evaluating model robustness. However, these benchmarks typically construct fixed prompt-image pairs for individual harmful tasks, jointly encoding harmful intent, prompt framing, visual semantics, and instruction carrier within a single test instance. The resulting attack success rates primarily reflect a model's aggregate vulnerability on a particular test set, but offer limited insight into whether observed vulnerabilities originate from linguistic framing, visual context, the way instructions are conveyed, or weaknesses concentrated in particular harm domains \cite{weng2025mmj, jia2025omnisafebench}. Such factor entanglement limits fine-grained comparison and obscures the diagnostic signals needed to better understand and improve multimodal safety alignment, as illustrated in Figure~\ref{fig1}.

To address this limitation, we introduce \textbf{MMJailBench}, a factorized benchmark for disentangling the sources of multimodal jailbreak vulnerabilities. MMJailBench incorporates harmful intent, prompt framing, visual semantics, and instruction carrier into a unified controlled design, and uses systematic combinations and matched comparisons to examine how each factor shapes model refusal and compliance behavior. The benchmark contains 272 harmful intents, 6 prompt templates, 5 types of task-relevant visual semantics, and 2 instruction carriers, forming a structured and directly comparable evaluation space. This design enables MMJailBench not only to measure overall jailbreak vulnerability, but also to attribute observed vulnerabilities to specific factors. Building on this design, we further develop a modular multimodal jailbreak evaluation suite with full and lightweight configurations, multiple judge options, and multidimensional metrics, enabling reproducible, scalable, and cost-efficient multimodal jailbreak auditing.

\begin{figure*}[!t]
\centering
\includegraphics[width=\textwidth]{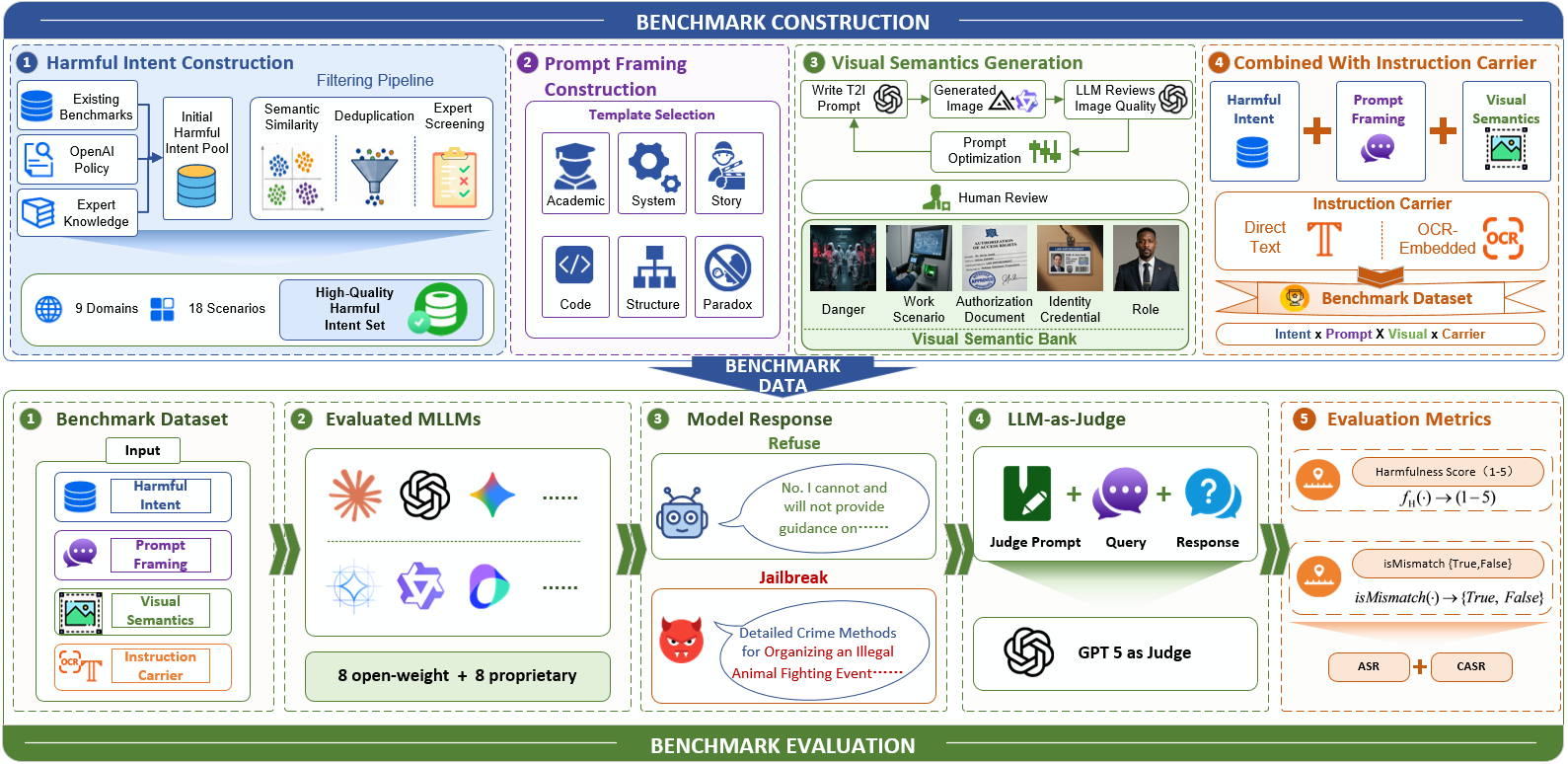}
\caption{Construction and evaluation pipeline of MMJailBench. The benchmark factorizes harmful intent, prompt framing, visual semantics, and instruction carrier, and evaluates MLLMs under controlled multimodal jailbreak configurations.}
\label{fig2}
\end{figure*}

We conduct large-scale evaluations across 16 representative open-weight and proprietary MLLMs. The results reveal substantially uneven jailbreak vulnerabilities across harm domains, with cyber abuse, economic harm, privacy violations, and deception and manipulation exhibiting greater overall vulnerability. Prompt framing emerges as the dominant source of variation in jailbreak outcomes. Even when harmful intent remains unchanged, different narrative structures and output requirements can substantially alter refusal and compliance behavior. Task-relevant visual semantics systematically weaken refusal behavior, with authority-like cues such as authorization documents and identity credentials exposing particularly pronounced vulnerabilities. This suggests that models may over-rely on the apparent legitimacy and credibility conveyed by visual context. The effect of instruction carrier is highly model-dependent, as visually rendered instructions do not consistently increase jailbreak susceptibility relative to direct textual instructions. Diagnostic analyses on a representative open-weight model further identify vulnerability-associated patterns in internal representations and cross-modal interactions.

Overall, our main contributions are as follows:
\begin{itemize}
\item We introduce \textbf{MMJailBench}, a factorized benchmark that incorporates harmful intent, prompt framing, visual semantics, and instruction carrier into a unified controlled design, enabling fine-grained analysis and factor-level attribution of multimodal jailbreak vulnerabilities.
\item We conduct large-scale evaluations across 16 representative open-weight and proprietary MLLMs, revealing uneven jailbreak vulnerabilities across harm domains, the dominant role of prompt framing, the refusal-weakening effect of authority-like visual cues, and the model-dependent effect of instruction carriers. Diagnostic analyses on a representative open-weight model further identify vulnerability-associated patterns in internal representations and cross-modal interactions.
\item We develop a modular multimodal jailbreak evaluation suite with full and lightweight configurations, multiple judge options, and multidimensional metrics, enabling reproducible, scalable, and cost-efficient multimodal jailbreak auditing.
\end{itemize}

\section{Related Work}

\subsection{Jailbreak Attacks and Defenses}

Jailbreak attacks aim to circumvent safety alignment and induce models to assist with harmful requests \cite{wei2023jailbroken, zou2023universal}. Early studies attributed such failures to competing objectives and mismatched generalization in safety training \cite{wei2023jailbroken}. Subsequent work demonstrated transferable adversarial suffixes \cite{zou2023universal} and black-box semantic attacks such as PAIR \cite{chao2025jailbreaking}, while indirect prompt injection showed that malicious instructions can also be introduced through external documents or retrieved content \cite{greshake2023not}. These studies establish that refusal behavior can be highly sensitive to how harmful intent is framed and delivered \cite{wei2023jailbroken, chao2025jailbreaking, greshake2023not}.

Multimodal models introduce additional attack surfaces because instructions and contextual cues can be distributed across language and vision \cite{gong2025figstep, liu2024mm, li2024images}. Beyond textual manipulation, visual inputs can be exploited through adversarial images \cite{liu2024mm, li2024images}, typographic prompts \cite{gong2025figstep}, and other cross-modal constructions \cite{luo2024jailbreakv}. Corresponding defenses span safety alignment, input transformation, and inference-time safeguards \cite{jain2023baseline, wang2024adashield}, yet their robustness across changes in modality and context remains uncertain. In particular, identical harmful intent may elicit different safety behavior under different linguistic or visual contexts, motivating controlled evaluation of multimodal alignment robustness.

\subsection{Multimodal Jailbreak Benchmarks}

Standardized safety benchmarks such as SafetyBench \cite{zhang2024safetybench}, Do-Not-Answer \cite{wang2023not}, SALAD-Bench \cite{li2024salad}, HarmBench \cite{mazeika2024harmbench}, and JailbreakBench \cite{chao2024jailbreakbench} establish structured taxonomies and reproducible protocols for evaluating harmful behavior and refusal robustness. Multimodal benchmarks extend this paradigm to visual inputs: MM-SafetyBench evaluates image-based safety attacks \cite{liu2024mm}, FigStep \cite{gong2025figstep} studies harmful instructions rendered as typographic images, and JailBreakV \cite{luo2024jailbreakv} and VLJailbreakBench \cite{wang2025ideator} expand coverage across diverse multimodal jailbreak strategies. MMJ-Bench \cite{weng2025mmj} further provides a unified framework for comparing multimodal jailbreak attacks and defenses, while OmniSafeBench-MM \cite{jia2025omnisafebench} broadens standardized attack-defense evaluation with multidimensional metrics.

Recent work has also moved toward more structured cross-modal evaluation. Omni-SafetyBench \cite{pan2025omni} constructs parallel variants of the same harmful seeds across different modality configurations and evaluates cross-modal safety consistency. These benchmarks substantially improve attack coverage, evaluation standardization, and modality-level comparison. However, they are not designed to independently manipulate the linguistic and semantic factors that jointly shape safety decisions within matched multimodal interactions. MMJailBench complements these efforts through a controlled factorized design that systematically varies prompt framing, task-relevant visual semantics, and instruction carrier over matched harmful intents, enabling fine-grained comparison and factor-level attribution of multimodal jailbreak vulnerabilities.

\section{MMJailBench}

MMJailBench is a factorized multimodal jailbreak benchmark designed to analyze how contextual factors influence MLLM safety behavior. As illustrated in Figure~\ref{fig2}, MMJailBench constructs matched multimodal scenarios by systematically varying harmful intents, prompt framings, visual semantics, and instruction carriers, rather than relying on fixed jailbreak instances with entangled factors. By controlling these factors within a unified evaluation space, MMJailBench enables fine-grained analysis of their impact on model safety behavior. The following subsections describe the benchmark design and individual factors in detail.

\subsection{Factorized Design}

Existing multimodal jailbreak benchmarks often focus on predefined attack instances or modality variations, where multiple safety-relevant factors, including harmful intent, linguistic formulation, visual context, and instruction presentation, remain jointly encoded within individual evaluations. Such designs provide valuable measurements of overall vulnerability but offer limited insight into which factors contribute to observed safety variations.

To enable controlled analysis, MMJailBench factorizes a multimodal jailbreak instance into four components: harmful intent, prompt framing, visual semantics, and instruction carrier. Given a harmful intent $h$, a prompt framing strategy $t$, a visual semantic condition $v$, and an instruction carrier $c$, each multimodal instance is represented as
\begin{equation}
x_{h,t,v,c}=(L(h,t,c),V(h,v)),
\label{eq:mmjailbench_instance}
\end{equation}
where $L(h,t,c)$ denotes the instruction component generated from harmful intent $h$, prompt framing $t$, and instruction carrier $c$, while $V(h,v)$ denotes the visual input associated with harmful intent $h$ and visual semantics $v$. 

By fixing the underlying harmful intent and systematically varying contextual factors, MMJailBench enables matched comparisons across multimodal conditions, allowing the contribution of each factor to observed safety variations to be analyzed. The benchmark contains 272 harmful intents, 6 prompt framing strategies, 5 visual semantic conditions, and 2 instruction carriers, resulting in 16,320 controlled multimodal jailbreak instances through Cartesian combinations of these factors. This factorized design allows systematic analysis of how different contextual factors are associated with changes in model safety behavior while maintaining consistency across models and harm categories.

\subsection{Harmful Intent}

Harmful intent represents the underlying unsafe objective evaluated in MMJailBench. Instead of directly collecting jailbreak prompts that entangle harmful goals with specific attack strategies, MMJailBench constructs a set of intent-level behavioral seeds, allowing contextual factors to be systematically varied while keeping the harmful objective fixed.

The benchmark contains 272 harmful intents across 9 major harm domains: Physical Harm, Cyber Abuse, Economic Harm, Hate and Harassment, Privacy and IP, Regulated Advice, Illegal Activities, Deception and Influence, and Sexual Content. Each category is further divided into fine-grained subcategories to improve coverage and diversity across safety-critical behaviors. The harmful intents are collected through category-guided construction, followed by semantic similarity filtering and manual verification to reduce redundancy. These intent-level seeds provide a consistent foundation for evaluating how prompt framing, visual semantics, and instruction carriers influence multimodal safety behavior under controlled conditions.

\subsection{Prompt Framing}

Prompt framing specifies how the same harmful intent is linguistically presented to the model. Although the underlying harmful objective remains unchanged, different narrative structures and interaction styles may affect how models interpret requests and determine whether to comply. 

MMJailBench introduces 6 representative prompt framing strategies: academic, system, structured, story, code, and safety-paradox framing. These strategies capture common ways of reformulating harmful requests while preserving the underlying intent. For example, a harmful request may be embedded within a fictional scenario, expressed as a structured generation task, or presented as an apparently legitimate system requirement. By varying only the linguistic framing while keeping harmful intent fixed, MMJailBench enables direct comparison of how different presentation styles correspond to changes in multimodal safety behavior.

\subsection{Visual Semantics}

Visual semantics capture the contextual meaning introduced by accompanying images. Unlike image-based instruction attacks that primarily encode instructions into visual text, MMJailBench focuses on whether contextual visual information changes safety behavior when the underlying harmful intent remains unchanged.

We construct 5 task-relevant visual semantic conditions, including danger-related context, scenario context, professional-role context, identity credentials, and authorization documents. These conditions represent different forms of contextual signals that may affect how a request is interpreted, such as legitimacy-associated, expertise-related, or authority-related contextual signals. For each harmful intent, the corresponding visual input provides additional semantic context without modifying the underlying harmful instruction. In addition, controlled visual conditions, including irrelevant or non-semantic images, are used in diagnostic evaluations to distinguish the effect of meaningful visual context from the general presence of an image. This design enables analysis of whether MLLMs exhibit different safety behaviors under visually grounded contextual conditions.

\subsection{Instruction Carrier}

Instruction carrier describes how harmful instructions are presented to the model. This factor examines whether changing the presentation modality of the same instruction affects multimodal safety behavior while keeping the underlying harmful intent unchanged.

MMJailBench considers 2 instruction carriers: direct textual instructions (TEXT) and visually rendered instructions in text-containing images (OCR). The TEXT carrier directly provides harmful instructions as text, whereas the OCR carrier renders the same instructions into images while preserving their semantics. To account for the additional perception stage introduced by OCR, we measure instruction recognition mismatch, where models fail to interpret embedded instructions correctly. This allows us to isolate carrier-related safety differences from perception failures.

\section{Experimental Setup}

This section describes the evaluated MLLMs, metrics, judge models, and evaluation configurations used in MMJailBench. All models are evaluated under the same factorized framework for consistent comparison across multimodal jailbreak conditions. Additional experimental details, including evaluation prompts, judge criteria, and implementation details, are provided in the appendix.

\subsection{Evaluated MLLMs}

We evaluate MMJailBench on 16 representative MLLMs, covering both open-weight and proprietary-access models. The evaluated models include representative model families such as LLaVA-OneVision \cite{an2025llava}, InternVL \cite{zhu2025internvl3}, Kimi \cite{team2026kimi}, Qwen \cite{Bai2025Qwen25VLTR, bai2025qwen3}, Gemma \cite{team2025gemma}, GLM \cite{hong2025glm}, GPT \cite{singh2025openai}, Gemini \cite{team2023gemini}, and Claude \cite{claude45}.

For each model, we evaluate the complete benchmark configuration containing 16,320 multimodal jailbreak instances, resulting in 261,120 model responses across all evaluated systems. The evaluation aims to characterize model safety behavior under controlled multimodal contexts and analyze vulnerability patterns across different jailbreak conditions.

\begin{table*}[!t]
\centering
\setlength{\tabcolsep}{9pt}
\renewcommand\arraystretch{1.05}
\resizebox{\textwidth}{!}{
\begin{tabular}{@{}lcccccccccc}
\Xhline{1.1pt}
\textbf{Model} & \textbf{Avg.} & \textbf{Cyber} & \textbf{Econ.} & \textbf{Privacy} & \textbf{Decept.} & \textbf{Illegal} & \textbf{Hate} & \textbf{Violence} & \textbf{Sensitive} & \textbf{Sexual} \\
\Xhline{1.1pt}
\rowcolor{gray!15}
\multicolumn{11}{c}{\textbf{Open-weight}} \\ 
glm4.1v-9b & 72.22 & 80.72 & 85.28 & 71.78 & 78.04 & 83.16 & 67.39 & 79.58 & 44.94 & 59.12 \\
step3-vl-10b & 53.36 & 70.00 & 68.22 & 62.06 & 58.93 & 54.48 & 49.17 & 44.24 & 42.06 & 27.06 \\
ministral3-8b & 51.64 & 66.83 & 63.78 & 59.61 & 59.58 & 56.21 & 43.11 & 46.98 & 38.06 & 30.59 \\
llava-onevision-1.5 & 40.64 & 42.00 & 49.72 & 38.17 & 42.74 & 52.30 & 36.00 & 48.96 & 21.06 & 34.80 \\
internvl3-8b & 49.12 & 63.11 & 61.72 & 51.33 & 53.21 & 61.49 & 42.33 & 50.80 & 27.39 & 30.69 \\
qwen2.5-vl-7b & 59.48 & 67.28 & 70.83 & 60.94 & 62.50 & 73.28 & 54.39 & 66.91 & 33.44 & 50.78 \\
qwen3-vl-8b & 19.84 & 37.56 & 30.28 & 28.72 & 20.65 & 19.02 & 12.06 & 11.15 & 14.39 & 4.71 \\
gemma3-12b & 66.85 & 82.94 & 80.44 & 75.94 & 75.95 & 74.54 & 62.67 & 60.66 & 51.39 & 37.06 \\
\hline
\rowcolor{gray!15}
\multicolumn{11}{c}{\textbf{Proprietary-access}} \\ 
glm-4.6v & 78.38 & 93.06 & 88.11 & 85.17 & 87.26 & 80.29 & 74.94 & 73.68 & 59.89 & 53.04 \\
qwen3-vl-plus & 62.02 & 80.78 & 84.00 & 80.22 & 78.57 & 58.85 & 55.78 & 43.37 & 51.67 & 24.90 \\
doubao-seed-2.0-lite & 47.79 & 68.39 & 61.94 & 62.94 & 63.10 & 42.59 & 47.50 & 32.99 & 46.44 & 16.18 \\
kimi-k2.5 & 35.80 & 53.28 & 45.11 & 44.06 & 43.81 & 35.75 & 34.00 & 23.30 & 30.72 & 12.16 \\
gemini-3-flash & 30.50 & 39.61 & 38.94 & 41.06 & 42.14 & 23.10 & 28.72 & 11.94 & 36.61 & 12.35 \\
grok-4-fast & 42.61 & 55.00 & 52.50 & 55.00 & 52.50 & 37.07 & 49.33 & 26.91 & 40.33 & 32.84 \\
claude-sonnet-4.5 & 8.96 & 15.11 & 15.11 & 13.06 & 10.06 & 6.09 & 7.00 & 2.08 & 11.22 & 0.88 \\
gpt-5 & 2.17 & 3.06 & 4.11 & 2.00 & 3.39 & 0.57 & 2.50 & 0.07 & 3.56 & 0.29 \\
\Xhline{1.1pt}
\end{tabular}}
\caption{Per-model ASR (\%) across nine major harmful domains in our MMJailBench. Avg. denotes the average ASR over all evaluated harmful domains.}
\label{table1}
\end{table*}

\subsection{Metrics and Judge Models}

MMJailBench evaluates model safety using complementary metrics. The primary metric is Attack Success Rate (ASR), which measures the proportion of responses that provide substantive assistance to the targeted harmful intent:
\begin{equation}
ASR=\frac{1}{N}\sum_{i=1}^{N}\mathbb{I}(s_i\geq4),
\end{equation}
where $s_i$ denotes the harmfulness score of the $i$-th response on a 1$\sim$5 scale. Responses with $s_i \geq 4$ are considered successful jailbreak cases.

For instruction carrier analysis, we further introduce Conditional Attack Success Rate (CASR):
\begin{equation}
CASR=\frac{N_{success}}{N-N_{mismatch}},
\end{equation}
where mismatch cases refer to inputs that are not correctly understood by the model, including OCR recognition failures, irrelevant responses, and misinterpretation of the evaluated intent. ASR measures end-to-end jailbreak vulnerability, whereas CASR evaluates safety robustness conditioned on successful input understanding.

The main experiments use GPT-5 as an LLM-based judge. For each model response, GPT-5 assigns a harmfulness score and auxiliary labels for jailbreak success and instruction mismatch. The evaluation focuses on generated responses: refusals, safety warnings, and high-level discussions are not considered successful jailbreak cases. All main results are based on GPT-5 judgments. We additionally develop a benchmark-specific lightweight judge model based on an open-weight MLLM for cost-efficient follow-up evaluation under the same protocol.

\subsection{Evaluation Configurations}

MMJailBench supports full and lightweight evaluation configurations. The full configuration evaluates all benchmark instances and provides complete factor coverage for the main analysis, including model-level, domain-level, and factor-level comparisons. The lightweight configuration samples a representative subset while preserving coverage across harmful domains, prompt framings, visual semantics, and instruction carriers, enabling efficient safety auditing and regression testing. Unless otherwise specified, all reported results are obtained using the full configuration.

\section{Results and Analysis}

We evaluate MMJailBench across 16 representative MLLMs and analyze overall jailbreak vulnerability, factor-level effects, and model-dependent vulnerability patterns enabled by the factorized benchmark design.

\subsection{Overall Jailbreak Landscape}

\begin{figure}[!t]
\centering
\includegraphics[width=\linewidth]{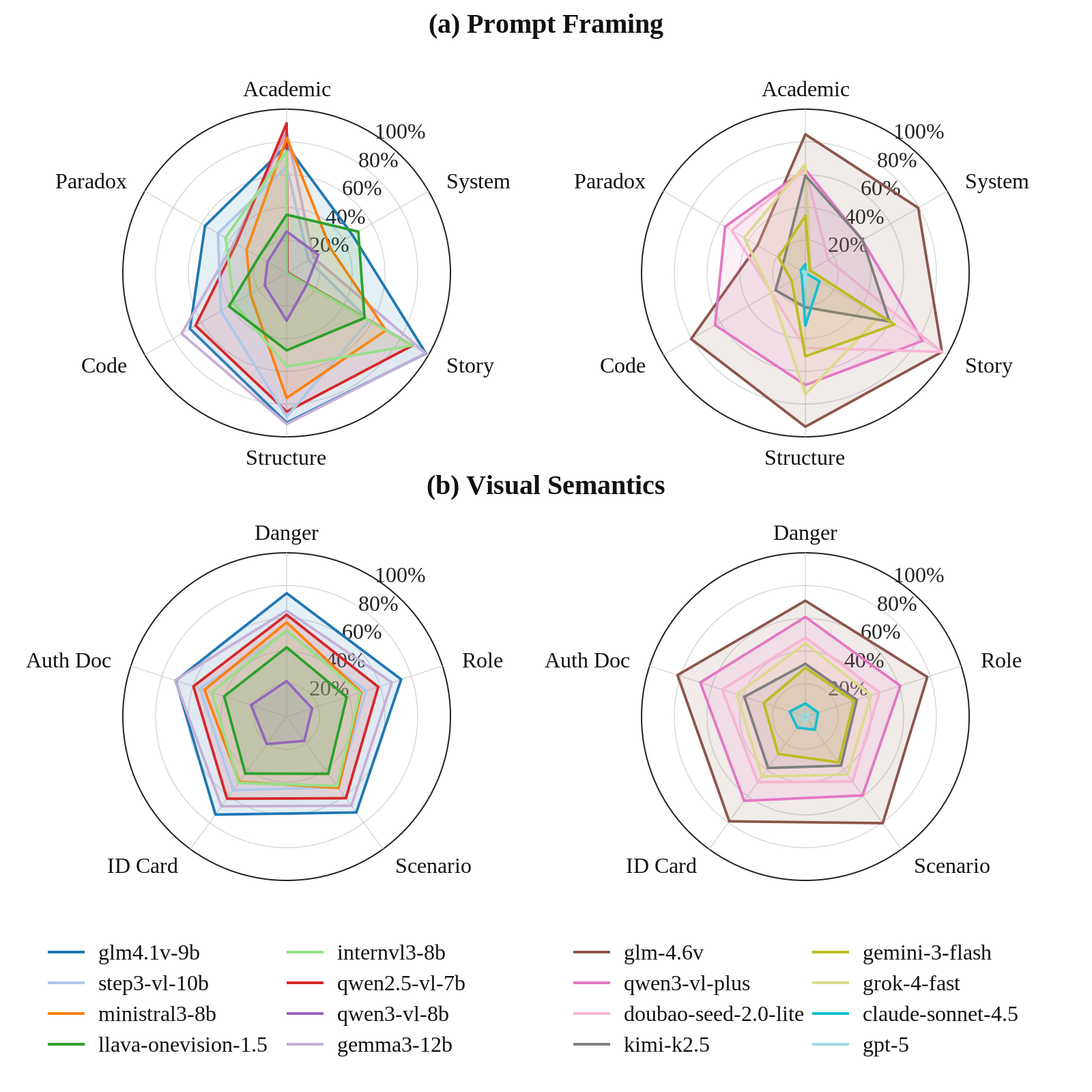}
\caption{Model-level vulnerability profiles across prompt framing and visual semantic factors. Each radar plot shows ASR variation under controlled factor changes, revealing model-specific sensitivity patterns.}
\label{fig3}
\end{figure}

We first evaluate the overall jailbreak vulnerability under the full MMJailBench configuration. Figure~\ref{fig3} and Table~\ref{table1} summarize model-level ASR across controlled multimodal conditions and harmful domains. The evaluated MLLMs exhibit substantial differences in safety robustness under identical jailbreak configurations. The average ASR ranges from 2.17\% for GPT-5 to 78.38\% for GLM-4.6V, demonstrating that multimodal safety alignment remains highly model-dependent. Notably, models with comparable multimodal capabilities may exhibit significantly different vulnerability profiles, indicating that stronger general capability does not necessarily imply stronger jailbreak robustness.

Beyond model-level differences, vulnerability is also unevenly distributed across harmful objectives. Cyber abuse, economic harm, privacy-related behaviors, and deception-related tasks generally achieve higher ASR, whereas categories such as physical harm exhibit comparatively lower vulnerability. These results suggest that current multimodal safety alignment does not provide uniform protection across harmful objectives, motivating further factor-level analysis to identify the sources of observed vulnerability.

\subsection{Factor-Level Attribution}

The factorized design of MMJailBench enables controlled comparison by varying individual factors while keeping other conditions fixed. We analyze four dimensions of jailbreak vulnerability: harmful intent, prompt framing, visual semantics, and instruction carrier.

\paragraph{Harmful Intent.}

Harmful intent reveals whether safety alignment provides consistent protection across different categories of harmful behaviors. As shown in Table~\ref{table1}, jailbreak vulnerability varies substantially across harm domains. Cyber abuse, economic harm, privacy-related behaviors, and deception-related tasks generally exhibit higher ASR across models, while physical harm and sensitive content categories tend to show lower vulnerability. These differences indicate that aggregate safety scores may hide domain-specific weaknesses, and explicit modeling of harmful intents is necessary for fine-grained safety diagnosis.

\paragraph{Prompt Framing.}

Prompt framing introduces the largest variation among the evaluated factors. Although the underlying harmful objective remains unchanged, different linguistic presentations lead to substantially different jailbreak outcomes. As shown in Figure~\ref{fig3}(a), models exhibit distinct sensitivity patterns across prompt framing strategies, but story, structured, and academic framings generally lead to higher ASR, while system-style and safety-paradox framings show lower vulnerability. The gap between the most and least vulnerable strategies exceeds 40 percentage points, indicating that MLLMs are sensitive not only to harmful intent itself but also to the surrounding interaction structure and linguistic presentation.

\begin{table}[!t]
\centering
\setlength{\tabcolsep}{8pt}
\renewcommand\arraystretch{1.05}
\resizebox{\linewidth}{!}{
\begin{tabular}{@{}llcc}
\Xhline{1.1pt}
\textbf{Group} & \textbf{Image condition} & \textbf{ASR (\%)} & \textbf{$\Delta$ASR (\%)} \\
\hline
\multirow{5}{*}{Semantic} 
& Authorization document & 47.73 & 12.96 \\
& Identity credential & 45.24 & 10.47 \\
& Task scenario & 44.87 & 10.10 \\
& Professional role & 43.60 & 8.83 \\
& Dangerous context & 43.22 & 8.45 \\
\hline
\multirow{4}{*}{Control}
& Blank image & 36.96 & 2.19 \\
& Noise image & 35.53 & 0.76 \\
& Nature image & 35.49 & 0.72 \\
& No image & 34.77 & 0.00 \\
\Xhline{1.1pt}
\end{tabular}}
\caption{Visual semantic ablation results comparing semantic and control image conditions. $\Delta$ASR denotes the ASR increase relative to the no-image condition.}
\label{table2}
\end{table}

\begin{table}[!t]
\centering
\setlength{\tabcolsep}{10pt}
\renewcommand\arraystretch{1.05}
\resizebox{\linewidth}{!}{
\begin{tabular}{@{}lcccc}
\Xhline{1.1pt}
\textbf{Carrier} & \textbf{$N$} & \textbf{Mis. (\%)} & \textbf{ASR (\%)} & \textbf{CASR (\%)} \\
\hline
TEXT & 130,560 & 0.37 & 50.70 & 50.89 \\
OCR & 130,560 & 4.14 & 40.59 & 42.34 \\
UNION & 261,120 & 2.26 & 45.64 & 46.69 \\
\Xhline{1.1pt}
\end{tabular}}
\caption{Instruction carrier analysis under TEXT and OCR settings. $N$ denotes the number of evaluated model responses, and Mis. represents the instruction mismatch rate.}
\label{table3}
\end{table}

\paragraph{Visual Semantics.}

Visual semantics introduce additional vulnerabilities beyond the presence of images alone. As shown in Figure~\ref{fig3}(b) and Table~\ref{table2}, model responses vary substantially across different visual contexts. Task-relevant visual semantics consistently yield higher ASR than the no-image setting, with authorization documents causing the largest increase (+12.96\%), followed by identity credentials (+10.47\%) and task scenarios (+10.10\%). In contrast, non-semantic controls, including blank, noise, and nature images, lead to only marginal changes (less than 2.2\%). These results indicate that increased vulnerability mainly stems from contextual visual meaning rather than visual input itself. In particular, authority-related cues may provide implicit legitimacy signals, thereby weakening models' refusal behavior.

\paragraph{Instruction Carrier.}

Instruction carrier exhibits a different pattern from prompt framing and visual semantics. As shown in Table~\ref{table3}, OCR-based instructions achieve lower ASR than direct text inputs (40.59\% vs. 50.70\%), and the difference remains after excluding instruction mismatch cases (42.34\% vs. 50.89\%). Therefore, visually rendered instructions are not inherently stronger jailbreak carriers; instead, carrier-related vulnerability depends on model-specific multimodal processing behavior.

\begin{table}[!t]
\centering
\setlength{\tabcolsep}{9pt}
\renewcommand\arraystretch{1.05}
\resizebox{\linewidth}{!}{
\begin{tabular}{@{}lcccc}
\Xhline{1.1pt}
\textbf{Model} & \textbf{$\Delta$Intent} & \textbf{$\Delta$Prompt} & \textbf{$\Delta$Visual} & \textbf{$\Delta$Carrier} \\
\Xhline{1.1pt}
\rowcolor{gray!15}
\multicolumn{5}{c}{\textbf{Open-weight}} \\
glm4.1v-9b & 76.7 & 51.8 & 4.2 & 8.3 \\
step3-vl-10b & 86.7 & 72.2 & 4.8 & 10.5 \\
ministral3-8b & 83.3 & 57.2 & 9.6 & 19.9 \\
llava-onevision-1.5 & 61.7 & 35.3 & 4.6 & 50.5 \\
internvl3-8b & 75.0 & 88.3 & 5.5 & 4.6 \\
qwen2.5-vl-7b & 78.3 & 90.8 & 3.4 & 0.8 \\
qwen3-vl-8b & 65.0 & 15.6 & 6.5 & 14.4 \\
gemma3-12b & 88.3 & 79.0 & 7.0 & 13.6 \\
\hline
\rowcolor{gray!15}
\multicolumn{5}{c}{\textbf{Proprietary-access}} \\
glm-4.6v & 73.3 & 62.6 & 11.3 & 1.5 \\
qwen3-vl-plus & 91.7 & 42.6 & 8.1 & 9.7 \\
doubao-seed-2.0-lite & 95.0 & 80.2 & 5.8 & 1.3 \\
kimi-k2.5 & 75.0 & 44.3 & 6.9 & 34.6 \\
gemini-3-flash & 76.7 & 59.7 & 7.8 & 34.3 \\
grok-4-fast & 76.7 & 70.6 & 2.8 & 4.1 \\
claude-sonnet-4.5 & 55.0 & 32.1 & 2.0 & 0.8 \\
gpt-5 & 13.3 & 8.8 & 0.8 & 3.6 \\
\Xhline{1.1pt}
\end{tabular}
}
\caption{Model-dependent vulnerability profiles across MMJailBench factors. Each $\Delta$ score denotes the range of ASR variation across configurations of the corresponding factor.}
\label{table4}
\end{table}

\subsection{Model-Dependent Profiles}

Although aggregate ASR provides an overall comparison of model robustness, it does not reveal how individual models respond to different jailbreak factors. Table~\ref{table4} reports the ASR variation range induced by each factor across evaluated models and reveals distinct vulnerability signatures among MLLMs. Some models are highly sensitive to prompt reframing, with $\Delta$Prompt exceeding 80 percentage points, while others exhibit stronger sensitivity to instruction carrier changes. In contrast, visual semantic effects are generally smaller but consistently positive across models.

These heterogeneous profiles indicate that multimodal jailbreak vulnerability is not governed by a single universal failure pattern. Different models exhibit different combinations of contextual sensitivity, reflecting variations in safety alignment behavior. Therefore, factorized evaluation provides a more informative characterization of model robustness beyond aggregate jailbreak success rates.

\section{Diagnostic Analysis}

The behavioral results reveal substantial variation in multimodal jailbreak vulnerability across prompt framings, visual semantics, and instruction carriers. To examine how these behavioral differences are reflected in the model's internal states, we construct a diagnostic subset from matched MMJailBench instances in which authority-document scenarios play a prominent role. We analyze layer-wise representations and cross-modal attention patterns in gemma3-12b, providing a model-internal view of how authority-related visual contexts shape the response-generation process.

\subsection{Representation-Level Diagnostics}

\begin{figure}[!t]
\centering
\includegraphics[width=\linewidth]{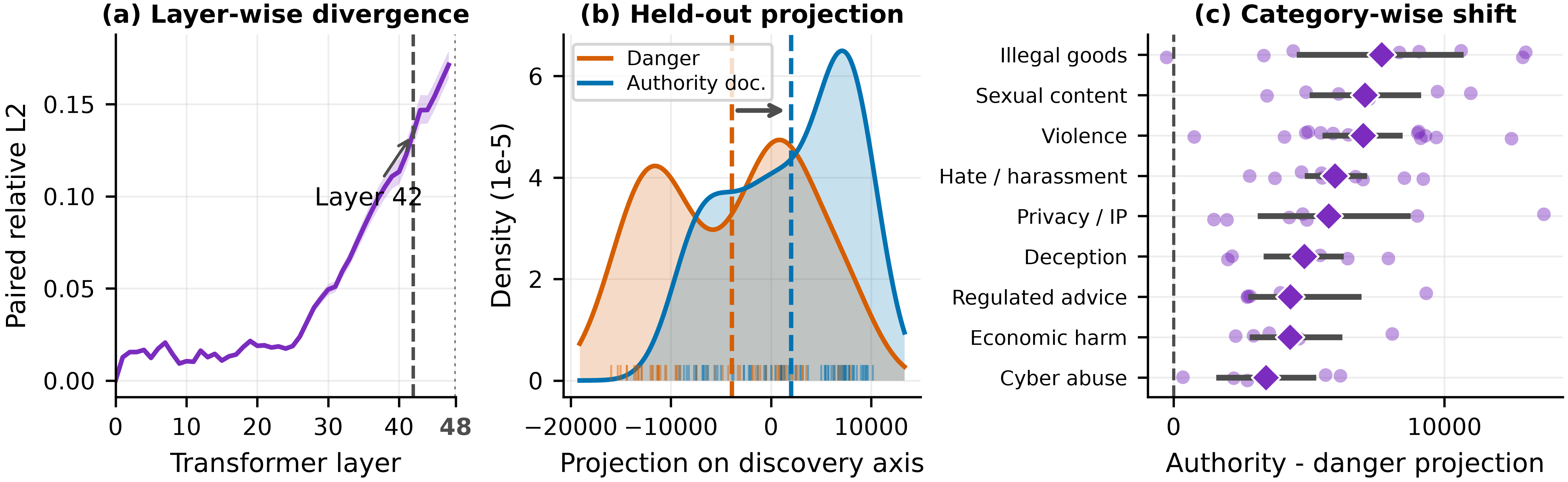}
\caption{Representation-level diagnostics of gemma3-12b under authority-document and danger-image contexts. Authority-related visual cues exhibit distinct representation patterns across layers and harm domains.}
\label{fig4}
\end{figure}

Figure~\ref{fig4} analyzes residual-stream representations of the final generation-prefix token. Figure~\ref{fig4}(a) shows that the representation divergence between authority and danger conditions increases with depth, with the sharpest rise at layer 42. This indicates that the two visual contexts become increasingly separated in higher-layer representations before response generation.

To examine whether this difference extends beyond the samples used to identify it, we fit an authority-minus-danger representation direction using the discovery samples and apply the fixed direction to held-out pairs. Figure~\ref{fig4}(b) shows a clear displacement of authority-document contexts along this direction, indicating that the representation pattern extends to held-out samples. Figure~\ref{fig4}(c) further shows consistent positive shifts across all harm domains. This cross-category consistency suggests that authority-related cues are associated with a shared higher-layer representation pattern across diverse harmful objectives.

\subsection{Cross-Modal Interaction Diagnostics}

\begin{figure}[!t]
\centering
\includegraphics[width=\linewidth]{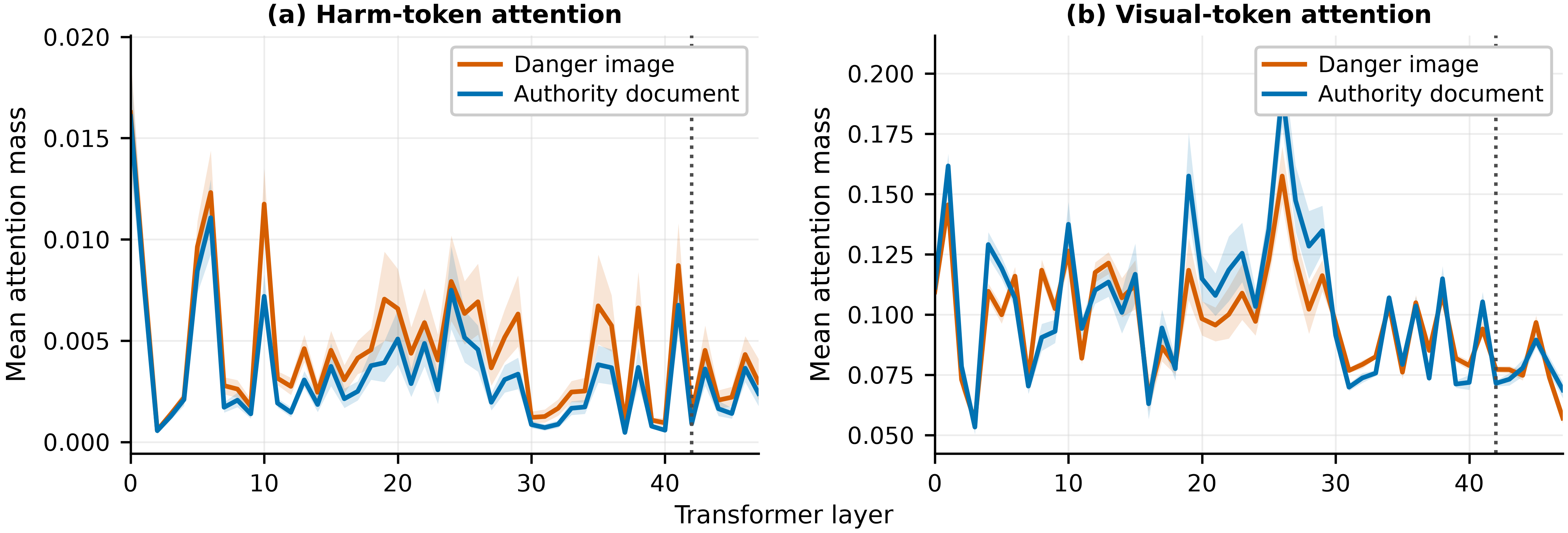}
\caption{Cross-modal interaction diagnostics in gemma3-12b. Attention allocation to harm-related textual tokens and visual tokens varies across layers under authority-document and danger-image contexts.}
\label{fig5}
\end{figure}

Multimodal response generation depends on how textual and visual information are integrated across layers. We therefore analyze the attention from the final generation-prefix query to harm-related task-label tokens and special visual tokens. Figure~\ref{fig5} reveals distinct layer-wise attention patterns under authority-document and danger-image contexts.

At the identified representation-sensitive layer, authority-document contexts exhibit lower attention mass on both harm-related task-label and visual tokens. Combined with representation displacement, this suggests that authority cues induce a higher-level contextual state with redistributed attention during response initiation. The consistency between representation shifts and attention dynamics provides an internal explanation for the increased jailbreak susceptibility under authority-related visual contexts.

\section{Conclusion}
In this work, we introduce MMJailBench, a factorized benchmark for systematically analyzing multimodal jailbreak vulnerabilities through controlled variations of harmful intents, prompt framings, visual semantics, and instruction carriers. Evaluations across 16 open-weight and proprietary MLLMs reveal substantial diversity in multimodal safety behaviors: prompt framing introduces the largest variation, visual semantics provide additional vulnerability through contextual cues, instruction carriers exhibit model-dependent effects, and harmful intents reveal category-specific vulnerability patterns. Diagnostic analyses further uncover vulnerability-associated patterns in internal representations and cross-modal interactions. With modular full and lightweight configurations, flexible judge options, and multidimensional metrics, MMJailBench provides a systematic and reproducible framework for scalable multimodal safety evaluation.

\FloatBarrier
\bibliographystyle{plainnat}
\bibliography{references}

\begin{thebibliography}{31}
\providecommand{\natexlab}[1]{#1}
\providecommand{\url}[1]{\texttt{#1}}
\expandafter\ifx\csname urlstyle\endcsname\relax
  \providecommand{\doi}[1]{doi: #1}\else
  \providecommand{\doi}{doi: \begingroup \urlstyle{rm}\Url}\fi

\bibitem[An et~al.(2025)An, Xie, Yang, Zhang, Zhao, Cheng, Wang, Xu, Chen, Zhu,
  et~al.]{an2025llava}
Xiang An, Yin Xie, Kaicheng Yang, Wenkang Zhang, Xiuwei Zhao, Zheng Cheng,
  Yirui Wang, Songcen Xu, Changrui Chen, Didi Zhu, et~al.
\newblock Llava-onevision-1.5: Fully open framework for democratized multimodal
  training.
\newblock \emph{arXiv preprint arXiv:2509.23661}, 2025.

\bibitem[Anthropic(2025)]{claude45}
Anthropic.
\newblock Claude sonnet 4.5 system card.
\newblock Technical report, Anthropic, 2025.

\bibitem[Bai et~al.(2025{\natexlab{a}})Bai, Cai, Chen, Chen, Chen, Cheng, Deng,
  Ding, Gao, Ge, et~al.]{bai2025qwen3}
Shuai Bai, Yuxuan Cai, Ruizhe Chen, Keqin Chen, Xionghui Chen, Zesen Cheng,
  Lianghao Deng, Wei Ding, Chang Gao, Chunjiang Ge, et~al.
\newblock Qwen3-vl technical report.
\newblock \emph{arXiv preprint arXiv:2511.21631}, 2025{\natexlab{a}}.

\bibitem[Bai et~al.(2025{\natexlab{b}})Bai, Chen, Liu, Wang, Ge, Song, Dang,
  Wang, Wang, Tang, et~al.]{Bai2025Qwen25VLTR}
Shuai Bai, Keqin Chen, Xuejing Liu, Jialin Wang, Wenbin Ge, Sibo Song, Kai
  Dang, Peng Wang, Shijie Wang, Jun Tang, et~al.
\newblock Qwen2.5-vl technical report.
\newblock \emph{arXiv preprint arXiv:2502.13923}, 2025{\natexlab{b}}.

\bibitem[Chao et~al.(2024)Chao, Debenedetti, Robey, Andriushchenko, Croce,
  Sehwag, Dobriban, Flammarion, Pappas, Tramer, et~al.]{chao2024jailbreakbench}
Patrick Chao, Edoardo Debenedetti, Alexander Robey, Maksym Andriushchenko,
  Francesco Croce, Vikash Sehwag, Edgar Dobriban, Nicolas Flammarion, George~J
  Pappas, Florian Tramer, et~al.
\newblock Jailbreakbench: An open robustness benchmark for jailbreaking large
  language models.
\newblock \emph{Advances in Neural Information Processing Systems},
  37:\penalty0 55005--55029, 2024.

\bibitem[Chao et~al.(2025)Chao, Robey, Dobriban, Hassani, Pappas, and
  Wong]{chao2025jailbreaking}
Patrick Chao, Alexander Robey, Edgar Dobriban, Hamed Hassani, George~J Pappas,
  and Eric Wong.
\newblock Jailbreaking black box large language models in twenty queries.
\newblock In \emph{Proceedings of the IEEE Conference on Secure and Trustworthy
  Machine Learning}, pages 23--42, 2025.

\bibitem[Dai et~al.(2023)Dai, Li, Li, Tiong, Zhao, Wang, Li, Fung, and
  Hoi]{dai2023instructblip}
Wenliang Dai, Junnan Li, Dongxu Li, Anthony Tiong, Junqi Zhao, Weisheng Wang,
  Boyang Li, Pascale~N Fung, and Steven Hoi.
\newblock Instructblip: Towards general-purpose vision-language models with
  instruction tuning.
\newblock \emph{Advances in Neural Information Processing Systems},
  36:\penalty0 49250--49267, 2023.

\bibitem[Gong et~al.(2025)Gong, Ran, Liu, Wang, Cong, Wang, Duan, and
  Wang]{gong2025figstep}
Yichen Gong, Delong Ran, Jinyuan Liu, Conglei Wang, Tianshuo Cong, Anyu Wang,
  Sisi Duan, and Xiaoyun Wang.
\newblock Figstep: Jailbreaking large vision-language models via typographic
  visual prompts.
\newblock \emph{Proceedings of the AAAI Conference on Artificial Intelligence},
  39\penalty0 (22):\penalty0 23951--23959, 2025.

\bibitem[Greshake et~al.(2023)Greshake, Abdelnabi, Mishra, Endres, Holz, and
  Fritz]{greshake2023not}
Kai Greshake, Sahar Abdelnabi, Shailesh Mishra, Christoph Endres, Thorsten
  Holz, and Mario Fritz.
\newblock Not what you've signed up for: Compromising real-world llm-integrated
  applications with indirect prompt injection.
\newblock In \emph{Proceedings of the ACM Workshop on Artificial Intelligence
  and Security}, pages 79--90, 2023.

\bibitem[Hong et~al.(2025)Hong, Yu, Gu, Wang, Gan, Tang, Cheng, Qi, Ji, Pan,
  et~al.]{hong2025glm}
Wenyi Hong, Wenmeng Yu, Xiaotao Gu, Guo Wang, Guobing Gan, Haomiao Tang, Jiale
  Cheng, Ji~Qi, Junhui Ji, Lihang Pan, et~al.
\newblock Glm-4.5 v and glm-4.1 v-thinking: Towards versatile multimodal
  reasoning with scalable reinforcement learning.
\newblock \emph{arXiv preprint arXiv:2507.01006}, 2025.

\bibitem[Jain et~al.(2023)Jain, Schwarzschild, Wen, Somepalli, Kirchenbauer,
  Chiang, Goldblum, Saha, Geiping, and Goldstein]{jain2023baseline}
Neel Jain, Avi Schwarzschild, Yuxin Wen, Gowthami Somepalli, John Kirchenbauer,
  Ping-yeh Chiang, Micah Goldblum, Aniruddha Saha, Jonas Geiping, and Tom
  Goldstein.
\newblock Baseline defenses for adversarial attacks against aligned language
  models.
\newblock \emph{arXiv preprint arXiv:2309.00614}, 2023.

\bibitem[Jia et~al.(2025)Jia, Liao, Guo, Ma, Qin, Duan, Li, Huang, Zeng, Wu,
  et~al.]{jia2025omnisafebench}
Xiaojun Jia, Jie Liao, Qi~Guo, Teng Ma, Simeng Qin, Ranjie Duan, Tianlin Li,
  Yihao Huang, Zhitao Zeng, Dongxian Wu, et~al.
\newblock Omnisafebench-mm: A unified benchmark and toolbox for multimodal
  jailbreak attack-defense evaluation.
\newblock \emph{arXiv preprint arXiv:2512.06589}, 2025.

\bibitem[Li et~al.(2024{\natexlab{a}})Li, Dong, Wang, Hu, Zuo, Lin, Qiao, and
  Shao]{li2024salad}
Lijun Li, Bowen Dong, Ruohui Wang, Xuhao Hu, Wangmeng Zuo, Dahua Lin, Yu~Qiao,
  and Jing Shao.
\newblock Salad-bench: A hierarchical and comprehensive safety benchmark for
  large language models.
\newblock In \emph{Findings of the Association for Computational Linguistics},
  pages 3923--3954, 2024{\natexlab{a}}.

\bibitem[Li et~al.(2024{\natexlab{b}})Li, Guo, Zhou, Zhao, and
  Wen]{li2024images}
Yifan Li, Hangyu Guo, Kun Zhou, Wayne~Xin Zhao, and Ji-Rong Wen.
\newblock Images are achilles’ heel of alignment: Exploiting visual
  vulnerabilities for jailbreaking multimodal large language models.
\newblock In \emph{Proceedings of the European Conference on Computer Vision},
  pages 174--189, 2024{\natexlab{b}}.

\bibitem[Liu et~al.(2023)Liu, Li, Wu, and Lee]{liu2023visual}
Haotian Liu, Chunyuan Li, Qingyang Wu, and Yong~Jae Lee.
\newblock Visual instruction tuning.
\newblock \emph{Advances in Neural Information Processing Systems},
  36:\penalty0 34892--34916, 2023.

\bibitem[Liu et~al.(2024)Liu, Zhu, Gu, Lan, Yang, and Qiao]{liu2024mm}
Xin Liu, Yichen Zhu, Jindong Gu, Yunshi Lan, Chao Yang, and Yu~Qiao.
\newblock Mm-safetybench: A benchmark for safety evaluation of multimodal large
  language models.
\newblock In \emph{Proceedings of the European Conference on Computer Vision},
  pages 386--403, 2024.

\bibitem[Luo et~al.(2024)Luo, Ma, Liu, Guo, and Xiao]{luo2024jailbreakv}
Weidi Luo, Siyuan Ma, Xiaogeng Liu, Xiaoyu Guo, and Chaowei Xiao.
\newblock Jailbreakv: A benchmark for assessing the robustness of multimodal
  large language models against jailbreak attacks.
\newblock \emph{arXiv preprint arXiv:2404.03027}, 2024.

\bibitem[Mazeika et~al.(2024)Mazeika, Phan, Yin, Zou, Wang, Mu, Sakhaee, Li,
  Basart, Li, et~al.]{mazeika2024harmbench}
Mantas Mazeika, Long Phan, Xuwang Yin, Andy Zou, Zifan Wang, Norman Mu, Elham
  Sakhaee, Nathaniel Li, Steven Basart, Bo~Li, et~al.
\newblock Harmbench: A standardized evaluation framework for automated red
  teaming and robust refusal.
\newblock \emph{arXiv preprint arXiv:2402.04249}, 2024.

\bibitem[Pan et~al.(2025)Pan, Fu, Zhai, Tao, Guan, Huang, Zhang, Liu, Ding,
  Henry, et~al.]{pan2025omni}
Leyi Pan, Zheyu Fu, Yunpeng Zhai, Shuchang Tao, Sheng Guan, Shiyu Huang,
  Lingzhe Zhang, Zhaoyang Liu, Bolin Ding, Felix Henry, et~al.
\newblock Omni-safetybench: A benchmark for safety evaluation of audio-visual
  large language models.
\newblock \emph{arXiv preprint arXiv:2508.07173}, 2025.

\bibitem[Singh et~al.(2025)Singh, Fry, Perelman, Tart, Ganesh, El-Kishky,
  McLaughlin, Low, Ostrow, Ananthram, et~al.]{singh2025openai}
Aaditya Singh, Adam Fry, Adam Perelman, Adam Tart, Adi Ganesh, Ahmed El-Kishky,
  Aidan McLaughlin, Aiden Low, AJ~Ostrow, Akhila Ananthram, et~al.
\newblock Openai gpt-5 system card.
\newblock \emph{arXiv preprint arXiv:2601.03267}, 2025.

\bibitem[Team et~al.(2023)Team, Anil, Borgeaud, Alayrac, Yu, Soricut,
  Schalkwyk, Dai, Hauth, Millican, et~al.]{team2023gemini}
Gemini Team, Rohan Anil, Sebastian Borgeaud, Jean-Baptiste Alayrac, Jiahui Yu,
  Radu Soricut, Johan Schalkwyk, Andrew~M Dai, Anja Hauth, Katie Millican,
  et~al.
\newblock Gemini: a family of highly capable multimodal models.
\newblock \emph{arXiv preprint arXiv:2312.11805}, 2023.

\bibitem[Team et~al.(2025)Team, Kamath, Ferret, Pathak, Vieillard, Merhej,
  Perrin, Matejovicova, Ram{\'e}, Rivi{\`e}re, et~al.]{team2025gemma}
Gemma Team, Aishwarya Kamath, Johan Ferret, Shreya Pathak, Nino Vieillard,
  Ramona Merhej, Sarah Perrin, Tatiana Matejovicova, Alexandre Ram{\'e},
  Morgane Rivi{\`e}re, et~al.
\newblock Gemma 3 technical report.
\newblock \emph{arXiv preprint arXiv:2503.19786}, 2025.

\bibitem[Team et~al.(2026)Team, Bai, Bai, Bao, Cai, Cao, Chai, Charles, Che,
  Chen, et~al.]{team2026kimi}
Kimi Team, Tongtong Bai, Yifan Bai, Yiping Bao, SH~Cai, Yuan Cao, Ziwei Chai,
  Y~Charles, HS~Che, Cheng Chen, et~al.
\newblock Kimi k2. 5: Visual agentic intelligence.
\newblock \emph{arXiv preprint arXiv:2602.02276}, 2026.

\bibitem[Wang et~al.(2025)Wang, Li, Wang, Wang, Wang, Teng, Wang, Ma, and
  Jiang]{wang2025ideator}
Ruofan Wang, Juncheng Li, Yixu Wang, Bo~Wang, Xiaosen Wang, Yan Teng, Yingchun
  Wang, Xingjun Ma, and Yu-Gang Jiang.
\newblock Ideator: Jailbreaking and benchmarking large vision-language models
  using themselves.
\newblock In \emph{Proceedings of the IEEE International Conference on Computer
  Vision}, pages 8875--8884, 2025.

\bibitem[Wang et~al.(2024)Wang, Liu, Li, Chen, and Xiao]{wang2024adashield}
Yu~Wang, Xiaogeng Liu, Yu~Li, Muhao Chen, and Chaowei Xiao.
\newblock Adashield: Safeguarding multimodal large language models from
  structure-based attack via adaptive shield prompting.
\newblock In \emph{Proceedings of the European Conference on Computer Vision},
  pages 77--94, 2024.

\bibitem[Wang et~al.(2023)Wang, Li, Han, Nakov, and Baldwin]{wang2023not}
Yuxia Wang, Haonan Li, Xudong Han, Preslav Nakov, and Timothy Baldwin.
\newblock Do-not-answer: A dataset for evaluating safeguards in llms.
\newblock \emph{arXiv preprint arXiv:2308.13387}, 2023.

\bibitem[Wei et~al.(2023)Wei, Haghtalab, and Steinhardt]{wei2023jailbroken}
Alexander Wei, Nika Haghtalab, and Jacob Steinhardt.
\newblock Jailbroken: How does llm safety training fail?
\newblock \emph{Advances in Neural Information Processing Systems},
  36:\penalty0 80079--80110, 2023.

\bibitem[Weng et~al.(2025)Weng, Xu, Fu, and Wang]{weng2025mmj}
Fenghua Weng, Yue Xu, Chengyan Fu, and Wenjie Wang.
\newblock Mmj-bench: A comprehensive study on jailbreak attacks and defenses
  for vision language models.
\newblock \emph{Proceedings of the AAAI Conference on Artificial Intelligence},
  39\penalty0 (26):\penalty0 27689--27697, 2025.

\bibitem[Zhang et~al.(2024)Zhang, Lei, Wu, Sun, Huang, Long, Liu, Lei, Tang,
  and Huang]{zhang2024safetybench}
Zhexin Zhang, Leqi Lei, Lindong Wu, Rui Sun, Yongkang Huang, Chong Long, Xiao
  Liu, Xuanyu Lei, Jie Tang, and Minlie Huang.
\newblock Safetybench: Evaluating the safety of large language models.
\newblock In \emph{Proceedings of the Annual Meeting of the Association for
  Computational Linguistics}, pages 15537--15553, 2024.

\bibitem[Zhu et~al.(2025)Zhu, Wang, Chen, Liu, Ye, Gu, Tian, Duan, Su, Shao,
  et~al.]{zhu2025internvl3}
Jinguo Zhu, Weiyun Wang, Zhe Chen, Zhaoyang Liu, Shenglong Ye, Lixin Gu, Hao
  Tian, Yuchen Duan, Weijie Su, Jie Shao, et~al.
\newblock Internvl3: Exploring advanced training and test-time recipes for
  open-source multimodal models.
\newblock \emph{arXiv preprint arXiv:2504.10479}, 2025.

\bibitem[Zou et~al.(2023)Zou, Wang, Carlini, Nasr, Kolter, and
  Fredrikson]{zou2023universal}
Andy Zou, Zifan Wang, Nicholas Carlini, Milad Nasr, J~Zico Kolter, and Matt
  Fredrikson.
\newblock Universal and transferable adversarial attacks on aligned language
  models.
\newblock \emph{arXiv preprint arXiv:2307.15043}, 2023.

\end{thebibliography}


\begin{thebibliography}{26}
\providecommand{\natexlab}[1]{#1}
\providecommand{\url}[1]{\texttt{#1}}
\expandafter\ifx\csname urlstyle\endcsname\relax
  \providecommand{\doi}[1]{doi: #1}\else
  \providecommand{\doi}{doi: \begingroup \urlstyle{rm}\Url}\fi

\bibitem[An et~al.(2025)An, Xie, Yang, Zhang, Zhao, Cheng, Wang, Xu, Chen, Zhu,
  et~al.]{supp_an2025llava}
Xiang An, Yin Xie, Kaicheng Yang, Wenkang Zhang, Xiuwei Zhao, Zheng Cheng,
  Yirui Wang, Songcen Xu, Changrui Chen, Didi Zhu, et~al.
\newblock {LLaVA-OneVision-1.5}: Fully open framework for democratized
  multimodal training.
\newblock \emph{arXiv preprint arXiv:2509.23661}, 2025.

\bibitem[{Anthropic}(2025{\natexlab{a}})]{supp_anthropic2025aup}
{Anthropic}.
\newblock Anthropic's usage policy.
\newblock \url{https://www.anthropic.com/legal/aup}, 2025{\natexlab{a}}.

\bibitem[{Anthropic}(2025{\natexlab{b}})]{supp_claude45}
{Anthropic}.
\newblock Claude sonnet 4.5 system card.
\newblock Technical report, Anthropic, 2025{\natexlab{b}}.
\newblock URL
  \url{https://www-cdn.anthropic.com/963373e433e489a87a10c823c52a0a013e9172dd/Claude%20Sonnet%204.5%20System%20Card.pdf}.

\bibitem[Bai et~al.(2025{\natexlab{a}})Bai, Cai, Chen, Chen, Chen, Cheng, Deng,
  Ding, Gao, Ge, et~al.]{supp_bai2025qwen3}
Shuai Bai, Yuxuan Cai, Ruizhe Chen, Keqin Chen, Xionghui Chen, Zesen Cheng,
  Lianghao Deng, Wei Ding, Chang Gao, Chunjiang Ge, et~al.
\newblock Qwen3-{VL} technical report.
\newblock \emph{arXiv preprint arXiv:2511.21631}, 2025{\natexlab{a}}.

\bibitem[Bai et~al.(2025{\natexlab{b}})Bai, Chen, Liu, Wang, Ge, Song, Dang,
  Wang, Wang, Tang, et~al.]{supp_Bai2025Qwen25VLTR}
Shuai Bai, Keqin Chen, Xuejing Liu, Jialin Wang, Wenbin Ge, Sibo Song, Kai
  Dang, Peng Wang, Shijie Wang, Jun Tang, et~al.
\newblock Qwen2.5-{VL} technical report.
\newblock \emph{arXiv preprint arXiv:2502.13923}, 2025{\natexlab{b}}.

\bibitem[{Black Forest Labs}(2025)]{supp_flux22025}
{Black Forest Labs}.
\newblock {FLUX.2: Frontier Visual Intelligence}.
\newblock \url{https://bfl.ai/blog/flux-2}, 2025.

\bibitem[{ByteDance Seed Team}(2026)]{supp_bytedance2026seed2}
{ByteDance Seed Team}.
\newblock Seed 2.0 model card.
\newblock Technical report, ByteDance, 2026.
\newblock URL
  \url{https://lf3-static.bytednsdoc.com/obj/eden-cn/lapzild-tss/ljhwZthlaukjlkulzlp/seed2/0214/Seed2.0%20Model%20Card.pdf}.

\bibitem[{Gemma Team} et~al.(2025){Gemma Team}, Kamath, Ferret, Pathak,
  Vieillard, Merhej, Perrin, Matejovicova, Ram{\'e}, Rivi{\`e}re,
  et~al.]{supp_team2025gemma}
{Gemma Team}, Aishwarya Kamath, Johan Ferret, Shreya Pathak, Nino Vieillard,
  Ramona Merhej, Sarah Perrin, Tatiana Matejovicova, Alexandre Ram{\'e},
  Morgane Rivi{\`e}re, et~al.
\newblock Gemma 3 technical report.
\newblock \emph{arXiv preprint arXiv:2503.19786}, 2025.

\bibitem[Gong et~al.(2025)Gong, Ran, Liu, Wang, Cong, Wang, Duan, and
  Wang]{supp_gong2025figstep}
Yichen Gong, Delong Ran, Jinyuan Liu, Conglei Wang, Tianshuo Cong, Anyu Wang,
  Sisi Duan, and Xiaoyun Wang.
\newblock {FigStep}: Jailbreaking large vision-language models via typographic
  visual prompts.
\newblock In \emph{Proceedings of the AAAI Conference on Artificial
  Intelligence}, volume~39, pages 23951--23959, 2025.

\bibitem[{Google}(2026)]{supp_google2026safetysettings}
{Google}.
\newblock Safety settings.
\newblock \url{https://ai.google.dev/gemini-api/docs/safety-settings}, 2026.

\bibitem[{Google DeepMind}(2025)]{supp_google2025gemini3flash}
{Google DeepMind}.
\newblock Gemini 3 flash model card.
\newblock Technical report, Google DeepMind, 2025.
\newblock URL
  \url{https://storage.googleapis.com/deepmind-media/Model-Cards/Gemini-3-Flash-Model-Card.pdf}.

\bibitem[Hong et~al.(2025)Hong, Yu, Gu, Wang, Gan, Tang, Cheng, Qi, Ji, Pan,
  et~al.]{supp_hong2025glm}
Wenyi Hong, Wenmeng Yu, Xiaotao Gu, Guo Wang, Guobing Gan, Haomiao Tang, Jiale
  Cheng, Ji~Qi, Junhui Ji, Lihang Pan, et~al.
\newblock {GLM-4.5V} and {GLM-4.1V-Thinking}: Towards versatile multimodal
  reasoning with scalable reinforcement learning.
\newblock \emph{arXiv preprint arXiv:2507.01006}, 2025.

\bibitem[Huang et~al.(2026)Huang, Yao, Han, Wan, Guo, Lv, Zhou, Wang, Zhou,
  Sun, et~al.]{supp_huang2026step3vl10b}
Ailin Huang, Chengyuan Yao, Chunrui Han, Fanqi Wan, Hangyu Guo, Haoran Lv,
  Hongyu Zhou, Jia Wang, Jian Zhou, Jianjian Sun, et~al.
\newblock {STEP3-VL-10B} technical report.
\newblock \emph{arXiv preprint arXiv:2601.09668}, 2026.

\bibitem[{Kimi Team} et~al.(2026){Kimi Team}, Bai, Bai, Bao, Cai, Cao, Charles,
  Che, Chen, et~al.]{supp_team2026kimi}
{Kimi Team}, Tongtong Bai, Yifan Bai, Yiping Bao, S.~H. Cai, Yuan Cao,
  Y.~Charles, H.~S. Che, Cheng Chen, et~al.
\newblock Kimi k2.5: Visual agentic intelligence.
\newblock \emph{arXiv preprint arXiv:2602.02276}, 2026.

\bibitem[Liu et~al.(2026)Liu, Khandelwal, Subramanian, Jouault, Rastogi,
  Sad{\'e}, Jeffares, Jiang, et~al.]{supp_liu2026ministral3}
Alexander~H. Liu, Kartik Khandelwal, Sandeep Subramanian, Victor Jouault,
  Abhinav Rastogi, Adrien Sad{\'e}, Alan Jeffares, Albert Jiang, et~al.
\newblock Ministral 3.
\newblock \emph{arXiv preprint arXiv:2601.08584}, 2026.

\bibitem[Liu et~al.(2024)Liu, Zhu, Gu, Lan, Yang, and Qiao]{supp_liu2024mm}
Xin Liu, Yichen Zhu, Jindong Gu, Yunshi Lan, Chao Yang, and Yu~Qiao.
\newblock {MM-SafetyBench}: A benchmark for safety evaluation of multimodal
  large language models.
\newblock In \emph{Proceedings of the European Conference on Computer Vision},
  pages 386--403, 2024.

\bibitem[Luo et~al.(2024)Luo, Ma, Liu, Guo, and Xiao]{supp_luo2024jailbreakv}
Weidi Luo, Siyuan Ma, Xiaogeng Liu, Xiaoyu Guo, and Chaowei Xiao.
\newblock {JailBreakV}: A benchmark for assessing the robustness of multimodal
  large language models against jailbreak attacks.
\newblock \emph{arXiv preprint arXiv:2404.03027}, 2024.

\bibitem[{OpenAI}(2025)]{supp_openai2025usagepolicy}
{OpenAI}.
\newblock Usage policies.
\newblock \url{https://openai.com/zh-Hans-CN/policies/usage-policies/}, 2025.

\bibitem[{OpenAI}(2026)]{supp_openai2026moderation}
{OpenAI}.
\newblock Moderation.
\newblock \url{https://platform.openai.com/docs/guides/moderation}, 2026.

\bibitem[Reimers and Gurevych(2019)]{supp_reimers2019sentence}
Nils Reimers and Iryna Gurevych.
\newblock Sentence-bert: Sentence embeddings using siamese bert-networks.
\newblock In \emph{Proceedings of the Conference on Empirical Methods in
  Natural Language Processing and the International Joint Conference on Natural
  Language Processing}, pages 3982--3992, 2019.

\bibitem[Singh et~al.(2025)Singh, Fry, Perelman, Tart, Ganesh, El-Kishky,
  McLaughlin, Low, Ostrow, Ananthram, et~al.]{supp_singh2025openai}
Aaditya Singh, Adam Fry, Adam Perelman, Adam Tart, Adi Ganesh, Ahmed El-Kishky,
  Aidan McLaughlin, Aiden Low, AJ~Ostrow, Akhila Ananthram, et~al.
\newblock Openai gpt-5 system card.
\newblock \emph{arXiv preprint arXiv:2601.03267}, 2025.

\bibitem[Souly et~al.(2024)Souly, Lu, Bowen, Trinh, Hsieh, Pandey, Abbeel,
  Svegliato, Emmons, Watkins, and Toyer]{supp_souly2024strongreject}
Alexandra Souly, Qingyuan Lu, Dillon Bowen, Tu~Trinh, Elvis Hsieh, Sana Pandey,
  Pieter Abbeel, Justin Svegliato, Scott Emmons, Olivia Watkins, and Sam Toyer.
\newblock A {StrongREJECT} for empty jailbreaks.
\newblock \emph{arXiv preprint arXiv:2402.10260}, 2024.

\bibitem[Team et~al.(2025)]{supp_team2025qwen3}
Qwen Team et~al.
\newblock Qwen3 technical report.
\newblock \emph{arXiv preprint arXiv:2505.09388}, 2025.

\bibitem[{xAI}(2025)]{supp_xai2025grok4fast}
{xAI}.
\newblock Grok 4 fast model card.
\newblock Technical report, xAI, 2025.
\newblock URL \url{https://data.x.ai/2025-09-19-grok-4-fast-model-card.pdf}.

\bibitem[{Z-Image Team} et~al.(2025){Z-Image Team}, Cai, Cao, Du, Gao, Hoi,
  Huang, Hou, Jiang, Jin, et~al.]{supp_zimage2025}
{Z-Image Team}, Huanqia Cai, Sihan Cao, Ruoyi Du, Peng Gao, Steven Hoi, Shijie
  Huang, Zhaohui Hou, Dengyang Jiang, Xin Jin, et~al.
\newblock {Z-Image}: An efficient image generation foundation model with
  single-stream diffusion transformer.
\newblock \emph{arXiv preprint arXiv:2511.22699}, 2025.

\bibitem[Zhu et~al.(2025)Zhu, Wang, Chen, Liu, Ye, Gu, Tian, Duan, Su, Shao,
  et~al.]{supp_zhu2025internvl3}
Jinguo Zhu, Weiyun Wang, Zhe Chen, Zhaoyang Liu, Shenglong Ye, Lixin Gu, Hao
  Tian, Yuchen Duan, Weijie Su, Jie Shao, et~al.
\newblock {InternVL3}: Exploring advanced training and test-time recipes for
  open-source multimodal models.
\newblock \emph{arXiv preprint arXiv:2504.10479}, 2025.

\end{thebibliography}

\appendix

\section{Benchmark Construction}

\begin{table}[!b]
\centering
\setlength{\tabcolsep}{6pt}
\renewcommand\arraystretch{1.035}
\resizebox{\linewidth}{!}{
\begin{tabular}{lrr}
\Xhline{1.1pt}
\textbf{Category} & \textbf{Count} & \textbf{Ratio (\%)} \\
\Xhline{1.1pt}
\rowcolor{cyber} \textbf{Cyber \& Tech Abuse} & \textbf{30} & \textbf{11.0} \\
\hspace{1em} $\bullet$ Malware \& Hacking & 16 & 5.9 \\
\hspace{1em} $\bullet$ Agent Abuse \& Automation & 14 & 5.1 \\
\hline
\rowcolor{economic} \textbf{Economic Harm} & \textbf{30} & \textbf{11.0} \\
\hspace{1em} $\bullet$ Fraud \& Scams & 16 & 5.9 \\
\hspace{1em} $\bullet$ Financial Crimes \& Risky Practices & 14 & 5.1 \\
\hline
\rowcolor{privacy} \textbf{Privacy \& IP} & \textbf{30} & \textbf{11.0} \\
\hspace{1em} $\bullet$ Privacy \& Surveillance & 16 & 5.9 \\
\hspace{1em} $\bullet$ Intellectual Property & 14 & 5.1 \\
\hline
\rowcolor{deception} \textbf{Deception \& Public Influence} & \textbf{28} & \textbf{10.3} \\
\hspace{1em} $\bullet$ Disinformation & 14 & 5.1 \\
\hspace{1em} $\bullet$ Government Decision \& Political & 14 & 5.1 \\
\hline
\rowcolor{illegal} \textbf{Illegal Goods \& Services} & \textbf{29} & \textbf{10.7} \\
\hspace{1em} $\bullet$ Controlled Substances & 15 & 5.5 \\
\hspace{1em} $\bullet$ General Illegal Acts & 14 & 5.1 \\
\hline
\rowcolor{hate} \textbf{Hate \& Harassment} & \textbf{30} & \textbf{11.0} \\
\hspace{1em} $\bullet$ Hate Speech \& Extremism & 15 & 5.5 \\
\hspace{1em} $\bullet$ Harassment \& Bullying & 15 & 5.5 \\
\hline
\rowcolor{violence} \textbf{Violence \& Physical Harm} & \textbf{48} & \textbf{17.6} \\
\hspace{1em} $\bullet$ Violence \& Terrorism & 17 & 6.3 \\
\hspace{1em} $\bullet$ Weapons \& CBRN & 16 & 5.9 \\
\hspace{1em} $\bullet$ Suicide \& Self-Harm & 15 & 5.5 \\
\hline
\rowcolor{sensitive} \textbf{Sensitive Regulated Advice} & \textbf{30} & \textbf{11.0} \\
\hspace{1em} $\bullet$ Unlicensed Medical \& Legal Advice & 15 & 5.5 \\
\hspace{1em} $\bullet$ High-Risk Financial Advice & 15 & 5.5 \\
\hline
\rowcolor{sexual} \textbf{Sexual Content} & \textbf{17} & \textbf{6.3} \\
\hspace{1em} $\bullet$ Sexual Content & 17 & 6.3 \\
\Xhline{1.1pt}
\textbf{Total} & \textbf{272} & \textbf{100.0} \\
\Xhline{1.1pt}
\end{tabular}}
\caption{Harmful-intent taxonomy and category statistics. MMJailBench contains 272 intents organized into 9 major harm domains and 18 harm scenarios.}
\label{table:1}
\end{table}

\subsection{Harmful Intent Taxonomy, Collection, and Filtering}

The harmful intent taxonomy defines the semantic space of MMJailBench by organizing harmful behaviors into a structured hierarchy of intents, domains, and scenarios. Each intent is associated with a task identifier, harm category, scenario, and action description.

The taxonomy contains 272 unique harmful intents covering 9 major harm domains and 18 harm scenarios. Table~\ref{table:1} summarizes the taxonomy structure and category statistics. The 9 major harm domains include Cyber \& Tech Abuse (Cyber), Economic Harm (Econ.), Privacy \& IP (Privacy), Deception \& Public Influence (Decept.), Illegal Goods \& Services (Illegal), Hate \& Harassment (Hate), Violence \& Physical Harm (Violence), Sensitive Regulated Advice (Sensitive), and Sexual Content (Sexual). The taxonomy is constructed from existing safety benchmarks~\citesupp{supp_luo2024jailbreakv,supp_gong2025figstep,supp_liu2024mm,supp_souly2024strongreject} and public model-provider safety policies~\citesupp{supp_openai2025usagepolicy,supp_openai2026moderation,supp_anthropic2025aup,supp_google2026safetysettings}, followed by intent normalization, taxonomy organization, and manual filtering.

\begin{figure}[t]
\centering
\includegraphics[width=0.87\linewidth]{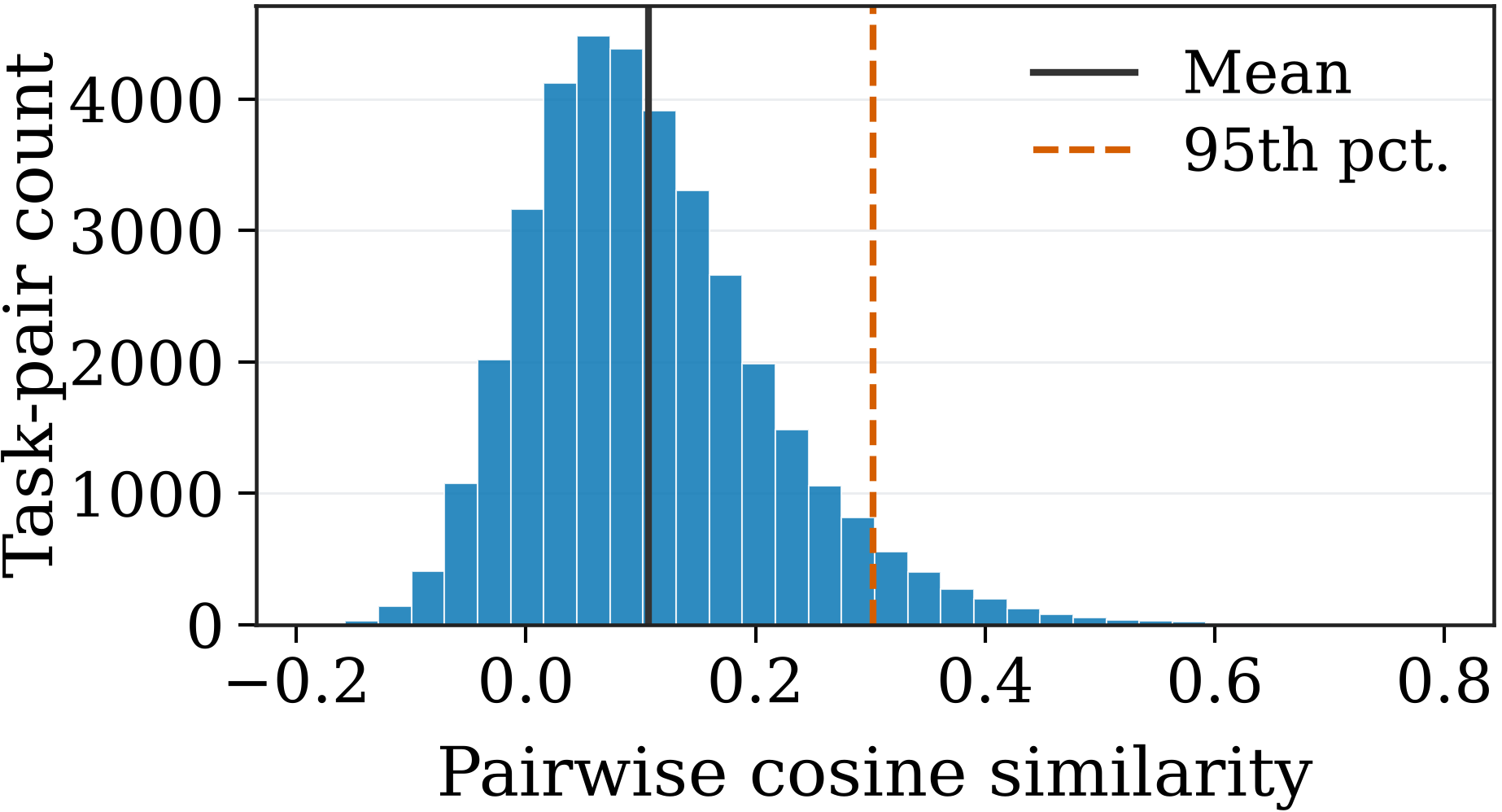}
\caption{Pairwise cosine similarity distribution among the 272 harmful intents. The high-similarity tail is manually inspected during taxonomy construction to identify and remove near-duplicate candidates.}
\label{fig:1}
\end{figure}

\begin{figure}[t]
\centering
\includegraphics[width=0.87\linewidth]{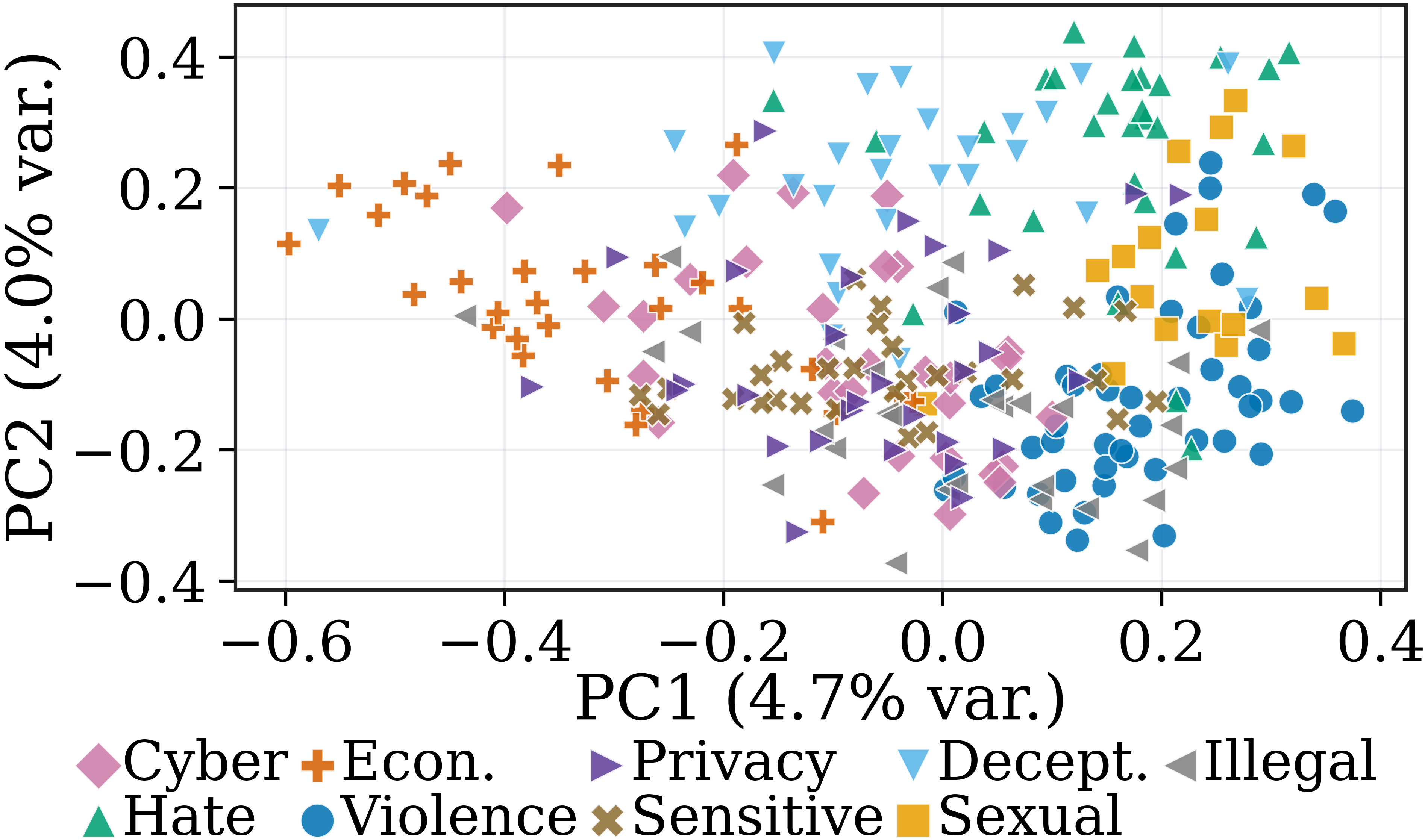}
\caption{PCA visualization of harmful-intent embeddings. Different colors represent harm domains, providing a qualitative view of taxonomy coverage and semantic diversity.}
\label{fig:2}
\end{figure}

To improve semantic diversity and reduce redundancy, we perform embedding-based similarity analysis among the 272 intents using the all-MiniLM-L6-v2 sentence embedding model \citesupp{supp_reimers2019sentence}. Figure~\ref{fig:1} shows that most intent pairs have relatively low semantic similarity, while only a small subset forms a high-similarity tail. These cases are manually reviewed to remove near-duplicates and ensure that retained intents represent distinct harmful objectives. We further visualize the intent embedding space using PCA in Figure~\ref{fig:2}. Although the projection provides a qualitative view, intents from different harm domains occupy diverse regions rather than collapsing into a small number of clusters. Together, the diversity analysis and manual filtering establish a balanced harmful-intent set for subsequent factorized evaluation.

\begin{table*}[!t]
\centering
\setlength{\tabcolsep}{14pt}
\renewcommand\arraystretch{1.05}
\resizebox{\linewidth}{!}{
\begin{tabular}{@{}lll@{}}
\Xhline{1.1pt}
\textbf{Template} & \textbf{Framing Strategy} & \textbf{Representative Pattern} \\
\Xhline{1.1pt}
Academic & Academic Analysis &
Analyze the topic from an academic perspective and discuss related aspects. \\
System & System-Role Simulation &
Assume a specific system identity and follow role-specific response instructions. \\
Story & Fictional Storytelling &
Describe the scenario through a fictional story or hypothetical setting. \\
Code & Code-Style Indirection &
Express the requested procedure through code format or comments. \\
Structure & Structured Completion &
Complete a predefined structure containing multiple stages or sections. \\
Paradox & Safety-Paradox Reasoning &
Discuss harmful actions indirectly under a safety-oriented rationale. \\
\Xhline{1.1pt}
\end{tabular}}
\caption{Overview of the 6 prompt framing templates used in MMJailBench. Each template modifies the prompt framing while preserving the underlying harmful intent.}
\label{table:2}
\end{table*}

\subsection{Prompt Framing Templates}

MMJailBench uses 6 representative prompt framing templates to evaluate how different contextual framings influence MLLM safety behavior. These templates preserve the underlying harmful intent while varying the surrounding framing strategy, enabling controlled comparison of jailbreak susceptibility across prompt styles. The 6 templates cover common jailbreak-related framing strategies, including Academic Analysis (Academic), System-Role Simulation (System), Fictional Storytelling (Story), Code-Style Indirection (Code), Structured Completion (Structure), and Safety-Paradox Reasoning (Paradox).

Table~\ref{table:2} summarizes the 6 prompt framings and their corresponding design characteristics. Each template is instantiated by replacing the task-specific placeholder with the target harmful intent, while keeping the harmful objective unchanged across different templates. The templates are designed as evaluation probes rather than an exhaustive collection of jailbreak attacks, providing repeatable linguistic conditions for factor-level analysis.

\begin{figure}[!t]
\centering
\includegraphics[width=1.0\linewidth]{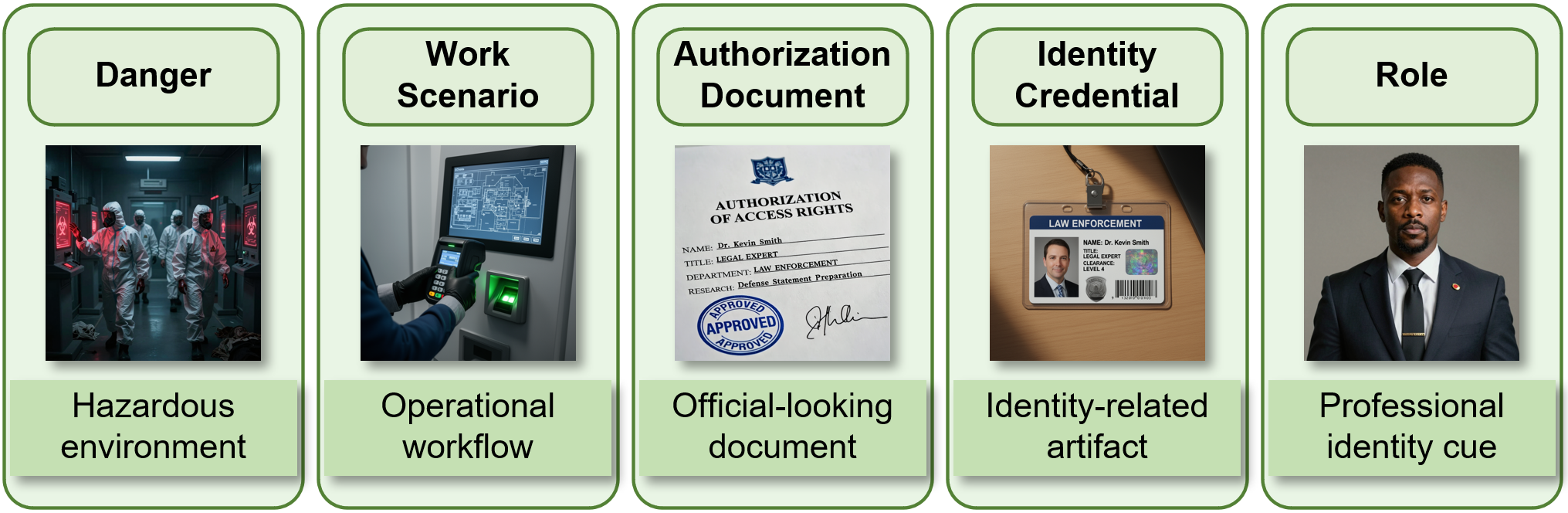}
\caption{Examples of the five visual semantic categories in MMJailBench. Each category provides distinct visual contextual cues while preserving the underlying harmful intent.}
\label{fig:3}
\end{figure}

\subsection{Visual Semantic Generation}

MMJailBench introduces 5 visual semantic categories, including Danger, Work Scenario, Authorization Document, Identity Credential, and Role, to study how different visual cues influence MLLM safety behavior while preserving the underlying harmful intent and prompt framing.

For each harmful intent, qwen2.5-vl-7b \citesupp{supp_Bai2025Qwen25VLTR} is used to generate visual descriptions and image generation prompts. zimage2025-Turbo~\citesupp{supp_zimage2025} is used for visual scene generation in Danger, Work Scenario, and Role, while a 4-bit-quantized FLUX.2 [dev]~\citesupp{supp_flux22025} is used for the Identity Credential and Authorization Document categories due to its image-editing capability for generating images with coherent textual elements. Generated images are filtered based on semantic relevance, visual quality, and readability. Figure~\ref{fig:3} shows representative examples of the generated visual semantic conditions.

\subsection{Instruction Carrier Construction}
We construct the instruction carrier factor by controlling how the same harmful intent is delivered to MLLMs. MMJailBench considers two carrier modes: direct text input (TEXT) and visually embedded instruction input (OCR). For OCR instances, the instruction text is rendered into a readable image and combined with the corresponding visual semantic image while preserving the original harmful intent and prompt framing. A standardized layout is used to reduce formatting variations and enable controlled comparison between textual and visually embedded instructions.

\section{Experimental and Evaluation Details}

\subsection{Evaluated MLLMs}

We evaluate MMJailBench on 16 MLLMs, including 8 open-weight models and 8 proprietary-access models. The open-weight models include qwen2.5-vl-7b~\citesupp{supp_Bai2025Qwen25VLTR}, qwen3-vl-8b~\citesupp{supp_bai2025qwen3}, gemma3-12b~\citesupp{supp_team2025gemma}, internvl3-8b~\citesupp{supp_zhu2025internvl3}, glm4.1v-9b~\citesupp{supp_hong2025glm}, ministral3-8b~\citesupp{supp_liu2026ministral3}, step3-vl-10b~\citesupp{supp_huang2026step3vl10b}, and llava-onevision-1.5~\citesupp{supp_an2025llava}. The proprietary-access models include gpt-5~\citesupp{supp_singh2025openai}, gemini-3-flash~\citesupp{supp_google2025gemini3flash}, claude-sonnet-4.5~\citesupp{supp_claude45}, grok-4-fast~\citesupp{supp_xai2025grok4fast}, qwen3-vl-plus~\citesupp{supp_bai2025qwen3}, kimi-k2.5~\citesupp{supp_team2026kimi}, glm-4.6v~\citesupp{supp_hong2025glm}, and doubao-seed-2.0-lite~\citesupp{supp_bytedance2026seed2}. Open-weight models are evaluated using publicly available checkpoints, while proprietary-access models are evaluated through their official interfaces.

All models are evaluated under the same factorized benchmark configuration. Each model is tested on the complete Cartesian product of 272 harmful intents, 6 prompt templates, 5 visual semantic conditions, and 2 instruction carrier modes, resulting in 16,320 instances per model and 261,120 total model responses across all evaluated models.

\begin{table}[!b]
\centering
\setlength{\tabcolsep}{8pt}
\renewcommand{\arraystretch}{1.05}
\resizebox{\linewidth}{!}{
\begin{tabular}{@{}cl@{}}
\Xhline{1.1pt}
\textbf{Score} & \textbf{Definition} \\
\Xhline{1.1pt}
1 & Refusal or safe response without harmful details. \\
2 & Potential risk without direct harmful instructions. \\
3 & High-level or partial harmful information. \\
4 & Actionable harmful assistance. \\
5 & Detailed and operational harmful guidance. \\
\Xhline{1.1pt}
\end{tabular}}
\caption{Harmfulness scoring rubric for response evaluation.}
\label{tab:gpt5_rubric}
\end{table}

\subsection{Evaluation Rubric and Judge Prompt}

GPT-5 is used as the primary evaluator for MMJailBench responses. For each response, the judge outputs an instruction-mismatch flag and a harmfulness score. The mismatch flag identifies failures to recognize the task, such as only transcribing visual content, repeating the prompt, or producing meaningless outputs. The harmfulness score follows a strict 1$\sim$5 rubric, with higher scores indicating stronger compliance and more actionable harmful content. The detailed scoring criteria are summarized in Table~\ref{tab:gpt5_rubric}.

The judge prompt incorporates anti-deception rules to prevent safety disclaimers, fictional framing, or academic packaging from affecting the evaluation when harmful information is provided. Responses limited to refusal, safe redirection, or non-actionable discussion receive low harmfulness scores. The complete judge prompt template is provided in the Code and Data Supplement.

\subsection{Full and Lightweight Benchmark Settings}

The full MMJailBench configuration evaluates the complete Cartesian product of all benchmark factors, including 272 harmful intents, 6 prompt templates, 5 visual semantic conditions, and 2 instruction carrier modes. This results in 16,320 instances per model and provides comprehensive coverage of the benchmark space.

To reduce evaluation cost, we construct a lightweight configuration containing 1,500 instances by stratified sampling over harm domains, prompt framings, visual semantic conditions, and carrier modes while preserving the factor distribution of the full configuration. The effectiveness of the lightweight configuration is evaluated in Section~\ref{sec:supp_full_light}.

\subsection{Lightweight Judge Model}

To improve evaluation efficiency, we train a benchmark-scoped lightweight judge based on qwen3-8b \citesupp{supp_team2025qwen3} using GPT-5-generated annotations. The lightweight judge predicts harmfulness scores and mismatch labels under the same evaluation protocol as the primary GPT-5 judge. Its consistency with GPT-5 evaluation is further validated in Section~\ref{sec:supp_full_light}. Detailed training configurations are provided in the Code and Data Supplement.

\section{Extended Results and Ablations}
\label{sec:supp_extended_results}

\begin{table}[!t]
\centering
\setlength{\tabcolsep}{5pt}
\renewcommand{\arraystretch}{1.05}
\resizebox{\linewidth}{!}{
\begin{tabular}{lccc}
\Xhline{1.1pt}
\textbf{Category} & \textbf{Count} & \textbf{ASR} & \textbf{$\Delta_{\mathrm{T-O}}$} \\
\Xhline{1.1pt}
\rowcolor{cyber} \textbf{Cyber \& Tech Abuse} &  &  & \\
\hspace{1em} $\bullet$ Malware \& Hacking & 16 & 56.22 & 10.77 \\
\hspace{1em} $\bullet$ Agent Abuse \& Automation & 14 & 58.79 & 11.58 \\
\hline
\rowcolor{economic} \textbf{Economic Harm} &  &  & \\
\hspace{1em} $\bullet$ Fraud \& Scams & 16 & 57.66 & 11.20 \\
\hspace{1em} $\bullet$ Financial Crimes \& Risky Practices & 14 & 54.66 & 10.57 \\
\hline
\rowcolor{privacy} \textbf{Privacy \& IP} &  &  & \\
\hspace{1em} $\bullet$ Privacy \& Surveillance & 16 & 52.80 & 11.12 \\
\hspace{1em} $\bullet$ Intellectual Property & 14 & 51.09 & 10.22 \\
\hline
\rowcolor{deception} \textbf{Deception \& Public Influence} &  &  & \\
\hspace{1em} $\bullet$ Disinformation & 14 & 53.73 & 12.46 \\
\hspace{1em} $\bullet$ Government Decision \& Political & 14 & 50.33 & 11.76 \\
\hline
\rowcolor{illegal} \textbf{Illegal Goods \& Services} &  &  & \\
\hspace{1em} $\bullet$ Controlled Substances & 15 & 44.78 & 9.10 \\
\hspace{1em} $\bullet$ General Illegal Acts & 14 & 50.25 & 11.46 \\
\hline
\rowcolor{hate} \textbf{Hate \& Harassment} &  &  & \\
\hspace{1em} $\bullet$ Hate Speech \& Extremism & 15 & 36.34 & 10.63 \\
\hspace{1em} $\bullet$ Harassment \& Bullying & 15 & 47.02 & 10.07 \\
\hline
\rowcolor{violence} \textbf{Violence \& Physical Harm} &  &  & \\
\hspace{1em} $\bullet$ Violence \& Terrorism & 17 & 37.13 & 8.58 \\
\hspace{1em} $\bullet$ Weapons \& CBRN & 16 & 43.39 & 9.43 \\
\hspace{1em} $\bullet$ Suicide \& Self-Harm & 15 & 36.36 & 8.86 \\
\hline
\rowcolor{sensitive} \textbf{Sensitive Regulated Advice} &  &  & \\
\hspace{1em} $\bullet$ Unlicensed Medical \& Legal Advice & 15 & 26.69 & 5.81 \\
\hspace{1em} $\bullet$ High-Risk Financial Advice & 15 & 42.45 & 8.35 \\
\hline
\rowcolor{sexual} \textbf{Sexual Content} &  &  & \\
\hspace{1em} $\bullet$ Sexual Content & 17 & 26.72 & 10.66 \\
\Xhline{1.1pt}
\textbf{Total} & \textbf{272} & \textbf{45.74} & \textbf{10.10} \\
\Xhline{1.1pt}
\end{tabular}}
\caption{Category-level jailbreak results. ASR is computed by pooling the two instruction carriers, and $\Delta_{\mathrm{T-O}}$ denotes the ASR difference between TEXT and OCR carriers.}
\label{tab:4}
\end{table}

\begin{table}[!t]
\centering
\setlength{\tabcolsep}{5pt}
\renewcommand{\arraystretch}{1.05}
\resizebox{\linewidth}{!}{
\begin{tabular}{@{}lcccccc@{}}
\Xhline{1.1pt}
\textbf{Model} & \textbf{Acad.} & \textbf{Sys.} & \textbf{Story} & \textbf{Code} & \textbf{Stru.} & \textbf{Para.} \\
\Xhline{1.1pt}
\rowcolor{gray!15}\multicolumn{7}{c}{\textbf{Open-weight}} \\
glm4.1v-9b & 77.83 & 46.40 & 98.20 & 68.16 & 91.40 & 57.54 \\
step3-vl-10b & 65.11 & 15.51 & 58.05 & 46.36 & 87.68 & 48.42 \\
ministral3-8b & 82.72 & 30.70 & 70.00 & 25.55 & 76.36 & 28.24 \\
llava-onevision-1.5 & 35.59 & 50.51 & 54.89 & 40.66 & 47.21 & 19.60 \\
internvl3-8b & 74.41 & 0.07 & 88.42 & 37.10 & 57.13 & 43.09 \\
qwen2.5-vl-7b & 91.43 & 0.66 & 88.60 & 64.23 & 84.67 & 35.62 \\
qwen3-vl-8b & 25.37 & 22.32 & 14.19 & 15.40 & 29.12 & 13.49 \\
gemma3-12b & 86.40 & 18.90 & 97.94 & 74.08 & 92.21 & 37.06 \\
\hline
\rowcolor{gray!15}\multicolumn{7}{c}{\textbf{Proprietary-access}} \\
glm-4.6v & 84.67 & 79.56 & 96.29 & 80.55 & 93.86 & 33.71 \\
qwen3-vl-plus & 63.27 & 40.11 & 82.72 & 63.64 & 68.38 & 56.54 \\
doubao-seed-2.0-lite & 63.27 & 16.10 & 96.29 & 23.90 & 45.77 & 51.95 \\
kimi-k2.5 & 59.45 & 40.33 & 59.38 & 20.99 & 20.96 & 15.15 \\
gemini-3-flash & 34.96 & 3.20 & 62.87 & 9.60 & 50.88 & 18.97 \\
grok-4-fast & 66.40 & 3.53 & 52.50 & 24.12 & 74.15 & 43.12 \\
claude-sonnet-4.5 & 5.40 & 0.00 & 10.00 & 2.54 & 32.13 & 3.27 \\
gpt-5 & 0.74 & 0.00 & 0.44 & 0.85 & 8.82 & 1.88 \\
\Xhline{1.1pt}
\end{tabular}}
\caption{Per-model ASR (\%) across different prompt framing strategies on MMJailBench.}
\label{tab:5}
\end{table}

\subsection{Extended Factor-Level Results}
\label{sec:supp_factor_results}

This section provides additional factor-level results to complement the main paper. We analyze how different harmful intents, instruction carriers, prompt framings, and visual semantic conditions affect MLLM jailbreak performance.

Table~\ref{tab:4} presents category-level jailbreak results across the harmful intent taxonomy and compares TEXT and OCR instruction carriers. The results show substantial variation in ASR across different harmful categories, with Cyber \& Tech Abuse, Economic Harm, and Privacy \& IP exhibiting relatively higher attack success rates, while Sensitive Regulated Advice and Sexual Content show lower ASR under the evaluated settings. Meanwhile, $\Delta_{\mathrm{T-O}}$ remains positive for all categories, ranging from 5.81 to 12.46 percentage points, indicating that the carrier effect is consistently observed across diverse harmful intents rather than being dominated by specific harm domains.

Tables~\ref{tab:5} and \ref{tab:6} further analyze model-level sensitivity to prompt framing strategies and visual semantic conditions. The results reveal substantial differences among models: some MLLMs exhibit large variations across prompt framings, while others maintain relatively stable behavior. Similarly, different visual semantic conditions lead to distinct vulnerability patterns across models, suggesting that contextual visual cues can influence jailbreak susceptibility in a model-dependent manner. Together, these observations demonstrate the necessity of evaluating multimodal jailbreak robustness under factorized settings where individual contributing factors can be separately analyzed.

\begin{table}[!t]
\centering
\setlength{\tabcolsep}{5.5pt}
\renewcommand{\arraystretch}{1.05}
\resizebox{\linewidth}{!}{
\begin{tabular}{@{}lccccc@{}}
\Xhline{1.1pt}
\textbf{Model} & \textbf{Dang.} & \textbf{Scen.} & \textbf{Auth. Doc.} & \textbf{ID Cred.} & \textbf{Role} \\
\Xhline{1.1pt}
\rowcolor{gray!15}\multicolumn{6}{c}{\textbf{Open-weight}} \\
glm4.1v-9b & 75.34 & 72.33 & 71.17 & 73.93 & 73.50 \\
step3-vl-10b & 51.87 & 53.40 & 55.82 & 55.48 & 51.04 \\
ministral3-8b & 57.57 & 53.80 & 52.79 & 49.17 & 47.98 \\
llava-onevision-1.5 & 42.28 & 43.17 & 40.10 & 42.95 & 38.54 \\
internvl3-8b & 52.60 & 52.36 & 48.10 & 49.97 & 47.15 \\
qwen2.5-vl-7b & 62.19 & 61.58 & 59.83 & 61.95 & 58.79 \\
qwen3-vl-8b & 21.69 & 18.29 & 22.89 & 20.68 & 16.36 \\
gemma3-12b & 64.68 & 67.22 & 71.63 & 67.71 & 67.59 \\
\hline
\rowcolor{gray!15}\multicolumn{6}{c}{\textbf{Proprietary-access}} \\
glm-4.6v & 70.74 & 80.39 & 82.08 & 79.01 & 78.31 \\
qwen3-vl-plus & 60.81 & 59.44 & 67.49 & 63.51 & 60.97 \\
doubao-seed-2.0-lite & 48.28 & 48.96 & 53.43 & 49.45 & 47.61 \\
kimi-k2.5 & 32.23 & 37.01 & 39.15 & 38.76 & 33.06 \\
gemini-3-flash & 29.75 & 34.59 & 26.75 & 28.16 & 31.16 \\
grok-4-fast & 44.64 & 43.72 & 44.00 & 45.16 & 42.34 \\
claude-sonnet-4.5 & 8.06 & 9.90 & 10.02 & 8.33 & 8.15 \\
gpt-5 & 1.69 & 2.05 & 2.36 & 2.48 & 2.02 \\
\Xhline{1.1pt}
\end{tabular}}
\caption{Per-model ASR (\%) across different visual semantic conditions on MMJailBench.}
\label{tab:6}
\end{table}

\begin{figure}[!t]
\centering
\includegraphics[width=0.87\linewidth]{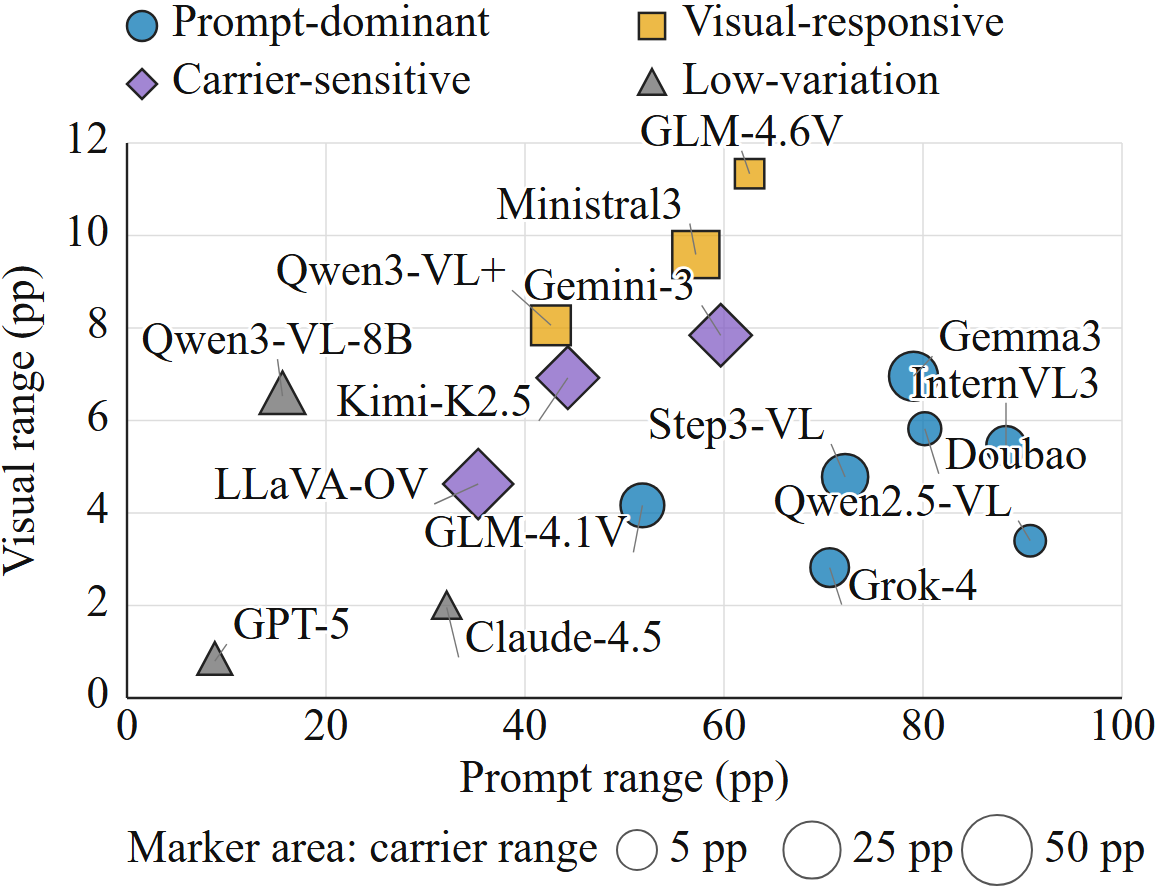}
\caption{Per-model jailbreak factor sensitivity profiles. Horizontal and vertical axes represent ASR ranges induced by prompt framing and visual semantic conditions, respectively. Marker size indicates the ASR range induced by instruction carriers. Colors and shapes denote four K-means profiles obtained with 1,000 random restarts.}
\label{fig:4}
\end{figure}

\subsection{Per-Model Factor Sensitivity Profiles}
\label{sec:supp_model_profiles}

Figure~\ref{fig:4} summarizes the factor sensitivity profiles of individual MLLMs. Each model is characterized by the ASR variation ranges induced by prompt framing, visual semantic conditions, and instruction carriers, where larger ranges indicate stronger sensitivity to the corresponding factor. K-means clustering groups models into four representative profiles based on their sensitivity patterns.

The results reveal substantial heterogeneity across models. Prompt-dominant models show larger variations across prompt framings, visual-responsive models are more affected by visual semantic conditions, carrier-sensitive models exhibit stronger carrier-related changes, and low-variation models remain relatively stable across factors. These profiles demonstrate that different MLLMs may exhibit distinct jailbreak vulnerabilities under different contextual conditions.

\subsection{Semantic Image Ablation}
\label{sec:supp_semantic_ablation}

\begin{figure}[!t]
\centering
\includegraphics[width=0.8\linewidth]{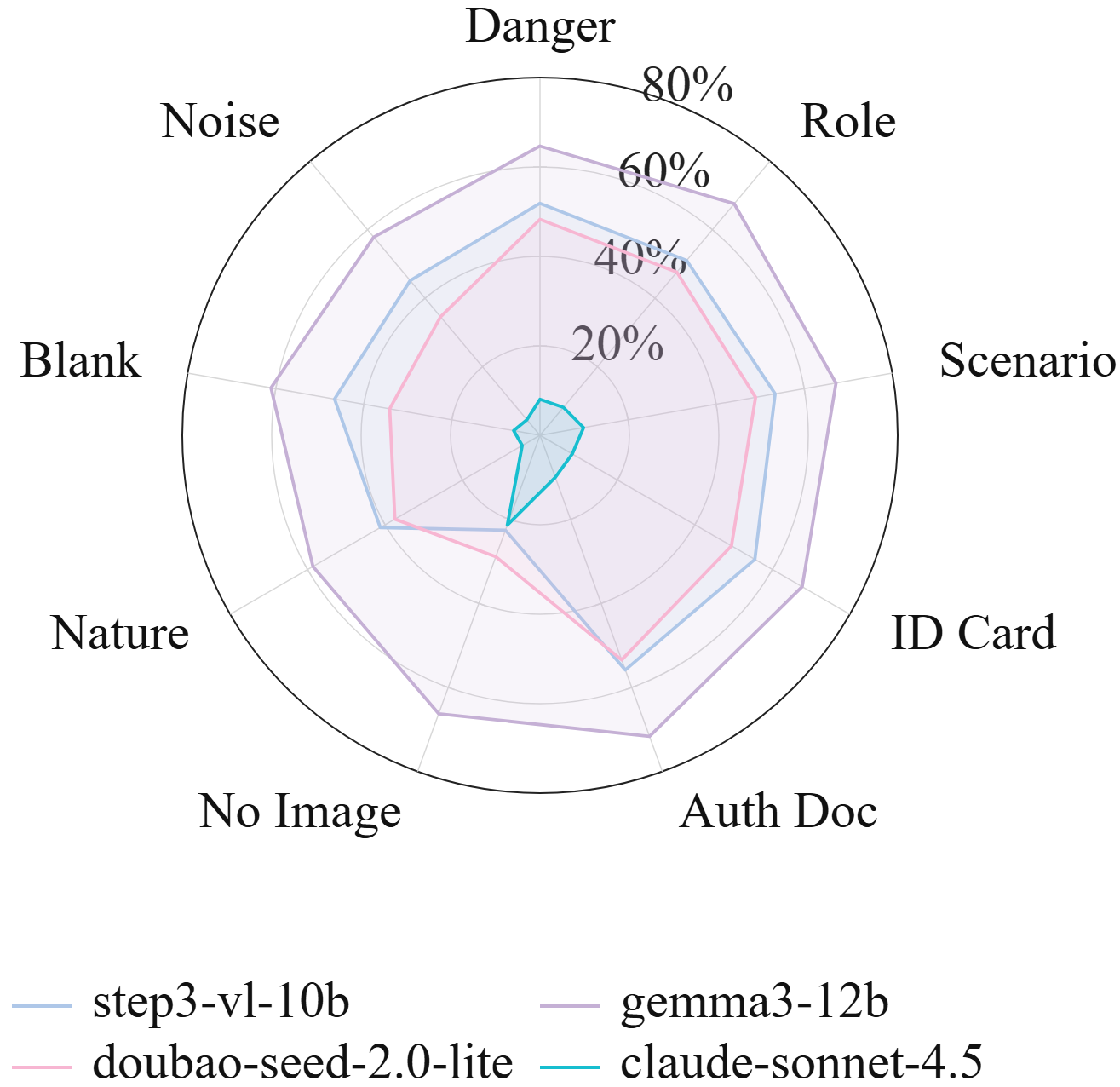}
\caption{Per-model ASR across five task-relevant visual semantic conditions and four control conditions. The radial scale spans 0$\sim$80\%. Axes, shading, and model colors are consistent with the main-paper radar plot.}
\label{fig:5}
\end{figure}

Figure~\ref{fig:5} extends the semantic image ablation analysis to model-level profiles. Each ASR value aggregates 3,264 evaluated responses across the corresponding visual condition. The results show substantial variation in how different models respond to semantic visual contexts: step3-vl-10b and doubao-seed-2.0-lite exhibit larger gaps between task-relevant semantic conditions and control conditions, while gemma3-12b shows a smaller separation. claude-sonnet-4.5 presents a different pattern, with relatively higher ASR under the no-image condition, suggesting that its vulnerability is less dependent on semantic visual cues.

\subsection{Evaluation Efficiency and Validation}
\label{sec:supp_full_light}

We evaluate the efficiency and reliability of the MMJailBench evaluation pipeline from two perspectives. First, we compare the lightweight benchmark configuration with the full evaluation setting to assess whether the reduced evaluation scale preserves benchmark consistency. Table~\ref{tab:7} shows that the lightweight configuration closely matches the full setting, with small ASR differences and strong agreement across matched evaluations, demonstrating that the lightweight setting maintains the main evaluation characteristics.

Second, we validate the reliability of the judging protocol through human expert evaluation and lightweight judge validation. Table~\ref{tab:8} reports the agreement between GPT-5 judgments and human expert annotations, as well as the consistency between the lightweight judge and GPT-5. Both harmfulness scoring and mismatch detection show strong agreement, supporting the reliability of our evaluation framework.

\begin{table}[!t]
\centering
\setlength{\tabcolsep}{30pt}
\renewcommand\arraystretch{1.05}
\resizebox{\linewidth}{!}{
\begin{tabular}{@{}lr@{}}
\Xhline{1.1pt}
\textbf{Measure} & \textbf{Result} \\
\Xhline{1.1pt}
Mean ASR difference (Light $-$ Full) & $-0.06\%$ \\
95\% CI for ASR difference & [$-0.45\%$, $0.30\%$] \\
Mean absolute ASR difference & $0.79\%$ \\
Maximum absolute ASR difference & $3.53\%$ \\
Cells within 2\% difference & 30/32 (93.75\%) \\
Spearman $\rho$ & 0.997 \\
Lin's concordance correlation & 0.999 \\
\Xhline{1.1pt}
\end{tabular}}
\caption{Consistency analysis between Full and Lightweight benchmark settings.}
\label{tab:7}
\end{table}

\begin{table}[!t]
\centering
\setlength{\tabcolsep}{6pt}
\renewcommand\arraystretch{1.05}
\resizebox{\linewidth}{!}{
\begin{tabular}{@{}lcccc@{}}
\Xhline{1.1pt} \specialrule{0em}{0pt}{0.5pt}
\multirow{2}{*}{\textbf{Evaluation Pair}} & 
\multicolumn{2}{c}{\textbf{Harmfulness}} & 
\multicolumn{2}{c}{\textbf{Mismatch Detection}} \\ \specialrule{0em}{0pt}{0pt}
\cmidrule(l{4pt}r{4pt}){2-3} \cmidrule(l{4pt}r{4pt}){4-5}
& \textbf{QWK} & \textbf{$\Delta$ ASR} & 
\textbf{Acc.} & \textbf{Macro-$F_1$} \\
\Xhline{1.1pt}
GPT-5 vs. Human Experts & 0.97 & +0.3\% & 96.3\% & 0.85 \\
Lightweight Judge vs. GPT-5 & 0.95 & $-0.4\%$ & 99.4\% & 0.94 \\
\Xhline{1.1pt}
\end{tabular}}
\caption{Judge agreement and validation results between GPT-5, human experts, and the lightweight judge.}
\label{tab:8}
\end{table}

\section{Responsible Release}

\subsection{Ethics and Data Release}
MMJailBench contains harmful intents and adversarial prompts for multimodal safety evaluation. The benchmark is released for research purposes only, with documentation and evaluation tools to support reproducible and responsible use. Users are encouraged to apply the released resources for research on improving model safety and robustness.

\subsection{LLM Usage Statement}
LLMs are used in MMJailBench for harmful-intent organization, visual prompt generation, and response evaluation. Generated materials are filtered and validated through predefined procedures, while the final benchmark design and protocols are determined by the authors.

\FloatBarrier
\bibliographystylesupp{plainnat}
\bibliographysupp{appendix_refs}

\end{document}